\documentclass[final,5p,times]{elsarticle}

\usepackage{amsmath, amssymb, mathtools}
\usepackage{amsthm}

\usepackage{tikz}
\usepackage{pgfplots}
\pgfplotsset{compat=1.18}
\usetikzlibrary{positioning}
\usetikzlibrary{arrows.meta,positioning,calc,shapes.geometric}

\usepackage{booktabs}
\usepackage{array}          
\usepackage{tabularx}       
\usepackage{longtable}      
\usepackage{makecell}       

\tikzset{
  >={Stealth[length=2.4mm,width=2mm]},
  arrow/.style={-Stealth,thick},
  dashedarrow/.style={dash pattern=on 2pt off 1.5pt,-Stealth,thick},
  block/.style={draw,rounded corners,minimum height=0.9cm,minimum width=3.6cm,align=center},
  smallblock/.style={draw,rounded corners,minimum height=0.8cm,minimum width=2.6cm,align=center},
  io/.style={draw,trapezium,trapezium left angle=70,trapezium right angle=110,minimum height=0.9cm,minimum width=3.6cm,align=center}
}

\usepackage{graphicx}
\usepackage{subcaption}
\usepackage{caption}

\usepackage{indentfirst}
\usepackage{balance}        

\usepackage{hyperref}
\hypersetup{
    colorlinks=true,
    linkcolor=blue,
    filecolor=magenta,
    urlcolor=cyan,
}

\usepackage{pifont}         
\usepackage{xspace}
\usepackage{listings}
\usepackage[table]{xcolor}
\rowcolors{3}{gray!6}{white}

\journal{---}

\begin{document}

\begin{frontmatter}

\title{Sliding-Mode Control Strategies for PMSM: Benchmarking and Comparative Simulation Study}

\author[1,2]{Mubarak Badamasi Aremu}
\author[1]{Abdullah Ajasa}
\author[1,2]{Ali Nasir} \ead{alinasir@kfupm.edu.sa}

\corref{Corresponding author: Ali Nasir (email: ali.nasir@kfupm.edu.sa)}

\affiliation[1]{organization={Control and Instrumentation Engineering Department, King Fahd University of Petroleum \& Minerals (KFUPM)},
            city={Dhahran}, 
            country={Saudi Arabia (SA)}.}

\affiliation[2]{organization={Interdisciplinary Research Center (IRC) for Intelligent Manufacturing and Robotics, KFUPM},
            country={SA}.}

\begin{abstract}
Permanent Magnet Synchronous Motors (PMSMs) are widely employed in high-performance drive systems owing to their high efficiency and power density. However, nonlinear dynamics, parameter uncertainties, and load disturbances complicate their control. Sliding-Mode Control (SMC) offers strong robustness but exists in numerous variants with unstandardized evaluation criteria. This paper presents a unified simulation benchmark and comparative analysis of six representative SMC techniques for PMSM speed regulation: conventional, integral, terminal, fractional-order, adaptive, and super-twisting. A standardized PMSM model, disturbance profile, and tuning protocol are adopted to ensure fair comparison across all methods. Performance is assessed through time-domain responses, integral error indices (ISE, IAE, ITSE, ITAE), and control-effort profiles, while also examining computational complexity and implementation feasibility. Results demonstrate that adaptive and higher-order SMCs, particularly the super-twisting and adaptive variants, achieve the most balanced trade-off between robustness, smoothness, and computational cost. The study provides a reproducible benchmarking framework, parameter-selection guidelines, and practical insights for designing efficient, low-chatter SMC-based PMSM drives suitable for real-time embedded implementation.
\end{abstract}


\end{frontmatter}

\section{Introduction} \label{sec: intro} 
Permanent Magnet Synchronous Motors (PMSMs) are among the most widely used electrical machines in advanced industrial systems due to their high efficiency, compact structure, and superior torque-to-inertia ratio \cite{podmiljvsak2024future}. They feature permanent magnets mounted either on the surface or embedded within the rotor (i.e., surface-mounted or interior permanent magnet synchronous motors, or PMSMs) \cite{lindh2009comparison}, producing a constant rotor flux without the need for excitation current. This design results in reduced losses and high power density \cite{pillay1989modeling}. PMSMs are extensively deployed in electric vehicles, robotics, precision manufacturing, renewable energy systems, and aerospace applications, where precise torque and speed control are critical \cite{karboua2023robust,zhao2023review,ozcciflikcci2024overview,azom2025recent,azom2025challenges}. Their combination of performance and compactness makes them ideal for applications with tight space and energy constraints. However, despite their growing adoption, PMSMs present significant control challenges that limit performance in real-world scenarios.

Strong nonlinearities, parameter uncertainties, and sensitivity to external disturbances characterize the dynamics of PMSMs. Factors such as load torque variations, magnetic saturation, temperature-dependent resistance drift, and back EMF coupling make the control task highly nontrivial \cite{nicola2020sensorless}. Moreover, high-performance PMSM control systems often require accurate position or speed feedback, which introduces sensor-related issues, including noise, cost, and reliability concerns \cite{gieras2018linear}. These challenges necessitate the use of robust, high-bandwidth control strategies that can ensure stability and performance across a wide range of operating conditions \cite{sakunthala2018review}.

Traditional control methods, such as Proportional-Integral-Derivative (PID) controllers and Field-Oriented Control (FOC), have been widely adopted in industrial Permanent Magnet Synchronous Motor (PMSM) drives due to their simplicity and ease of implementation \cite{zheng2011design}. In particular, Proportional-Integral (PI) controllers are commonly used for speed loop regulation because of their intuitive tuning and low computational overhead \cite{zhang2011direct}. However, their performance deteriorates significantly in the presence of nonlinearities, parameter uncertainties, and external disturbances. Notable limitations include excessive overshoot, poor robustness, strong dependency on accurate system modeling, and limited disturbance rejection capabilities. The increasing demand for high-performance PMSM drives, characterized by fast dynamic response, precise tracking, and resilience to system uncertainties, has driven the development of more advanced control strategies \cite{sakunthala2018review}. One persistent challenge in PMSM control is the requirement for rotor position information, which is typically obtained via position or speed sensors such as magnetic resolvers or optical encoders. While effective, these sensors introduce issues such as increased cost, sensitivity to noise, reduced reliability in harsh environments, and added system complexity.

Moreover, torque regulation in PMSMs is typically achieved by controlling the stator (armature) current, as electromagnetic torque is directly proportional to it \cite{zhong1999direct,frikha2023multiphase}. This has motivated the exploration of nonlinear and robust control techniques that can cope with the motor’s nonlinear behavior and sensitivity to operating conditions. Several approaches have been proposed in recent literature, including adaptive backstepping control for PMSMs \cite{wang2019adaptive,li2019model}, model predictive control (MPC) strategies for current loop optimization \cite{zhang2017performance,djouadi2024improved}, and intelligent controllers based on neural networks and fuzzy logic \cite{sakunthala2017study,nouaoui2024speed,mohajerani2024neural}. Despite these developments, achieving optimal control performance under real-world conditions remains challenging for conventional linear and gain-scheduled controllers due to the PMSM’s strong nonlinearities, coupled dynamics, and external disturbances. This has led to growing interest in SMC, which offers inherent robustness, disturbance rejection, and real-time adaptability to system uncertainties \cite{weijie2014sliding}.

SMC is a widely studied nonlinear control strategy known for its inherent robustness against system uncertainties and external disturbances \cite{gambhire2021review}. At its core, SMC operates by enforcing system trajectories onto a predefined sliding surface, where the closed-loop dynamics become insensitive to matched uncertainties. This is achieved through a discontinuous control law that drives the system to the sliding surface and maintains it there, ensuring fast convergence and strong disturbance rejection \cite{mohd2019robust}. Originally developed within the framework of variable structure systems in the mid-20th century, SMC has since been applied across diverse domains, including robotics, industrial automation, electric machinery, and aerospace systems. Its relevance to PMSM control stems from its ability to handle the nonlinear, coupled, and parameter-sensitive nature of the motor dynamics with high precision and resilience. Despite these advantages, classical SMC suffers from a significant drawback: chattering, which manifests as high-frequency oscillations in the control signal. This phenomenon, caused by the discontinuous nature of the control input, can induce mechanical wear, excite unmodeled dynamics, and impair tracking performance \cite{chen2019precision}. To mitigate this issue, numerous enhancements to the basic SMC framework have been proposed. Most SMC designs rely on Lyapunov-based asymptotic stability analysis, often utilizing linear switching manifolds to guarantee convergence. Various Lyapunov functions have been formulated to support different system structures and sliding surface designs \cite{luo2016novel, mohd2019robust}. To further improve system performance and reduce chattering, a wide range of advanced SMC variants have been introduced in recent years. These include higher-order SMC, terminal SMC, adaptive SMC, fractional-order SMC, and integral SMC, each aiming to reduce chattering and enhance smoothness, convergence speed, and implementation feasibility while preserving the robust nature of the earlier sliding mode control.

The earlier work \cite{ajasa2025} established a detailed taxonomy and critical
review of Sliding-Mode Control (SMC) techniques applied to Permanent-Magnet Synchronous Motor (PMSM) drives.  That analysis revealed that despite
significant conceptual diversity spanning conventional, integral, terminal,
fractional-order, adaptive, and higher-order super-twisting structures, the majority
of publications report performance improvements under disparate assumptions,
operating points, and evaluation indices.  As a result, the relative merits of each
SMC variants remain difficult to assess quantitatively.  A unified and fair comparison
under identical system conditions, it is therefore essential to bridge the gap between
literature theory and practical controller selection.

Despite extensive literature on SMC for PMSM drives, comparative validation across different variants is often hindered by inconsistent system models, load conditions, and evaluation metrics. Reported results are typically case-specific, making it difficult to quantify the relative advantages of each method in a unified setting. To bridge this gap, this work develops a standardized benchmarking environment that integrates modeling, control design, parameter tuning, and performance evaluation for six representative SMC strategies. The resulting framework enables objective comparison, provides implementation guidelines, and identifies computational and design trade-offs relevant to real-time embedded PMSM control.

Accordingly, this study provides a comprehensive simulation-based benchmark of six representative SMC variants for PMSM speed regulation. The objective is not merely to reproduce reported results but to analyze how intrinsic design features, such as the order of differentiation, adaptation law, and reaching dynamics, govern the trade-off between robustness, smoothness, and computational cost.  The six controllers considered are Conventional SMC (CSMC), Integral SMC (ISMC), Terminal SMC (TSMC), Fractional-Order SMC (FOSMC), Adaptive SMC (ASMC),
and Super-Twisting SMC (STSMC).  Each controller is implemented within a standard PMSM
plant model and evaluated under identical disturbance scenarios, ensuring a fair
basis for comparison.

Quantitative performance is measured using standard integral indices (ISE, IAE,
ITSE, ITAE) as well as qualitative assessment of chattering amplitude, settling time,
and disturbance-rejection behavior.  The comparative analysis not only benchmarks
these controllers but also elucidates the underlying mechanisms responsible for their
differences.  For example, STSMC’s continuous second-order dynamics minimize equivalent
control ripple, whereas ASMC’s gain-adaptation law mitigates chattering without
sacrificing transient speed.  Such insights provide actionable guidelines for
selecting or designing SMC schemes tailored to real-time PMSM applications. 

The main contributions of this work are summarized as follows:
\begin{enumerate}
    \item Establishment of a unified PMSM simulation and benchmarking framework enabling consistent evaluation of SMC strategies under identical plant parameters, load disturbances, and conditions.
    \item Six representative controllers (CSMC, ISMC, TSMC, FOSMC, ASMC, STSMC) are comparatively analyzed through standardized performance indices and computational benchmarks.
    \item Practical insights and design trade-offs are drawn to guide controller selection for high-performance PMSM drives.
    \item Formulation of a systematic parameter-tuning guideline for each controller based on Lyapunov stability analysis and empirical optimization.
    \item Benchmarking of computational complexity, memory cost, and implementation feasibility to inform real-time embedded control design.
    \item Delivery of quantitative insights and design recommendations for selecting robust, low-chatter controllers in high-performance PMSM drives.
\end{enumerate}

This unified benchmark serves as a reference framework for researchers developing next-generation adaptive, fractional-order, and learning-enhanced SMC strategies for PMSM systems. The rest of this paper is organized as follows. Section II introduces the configuration and properties of PMSM. Section III presents the unified PMSM model and control framework. Section IV discusses the sliding mode control (SMC) framework. Section V discusses the various SMC control variant architectures, and Section VI provides an in-depth analysis. At the same time, Section VII concludes the study with practical recommendations and future research directions.

\section{PMSM System and Objectives}

\subsection{Modeling Assumptions and Operating Conditions}
The PMSM model assumes sinusoidal stator windings, negligible magnetic saturation and core losses, balanced three-phase currents, and constant parameters within the tested temperature range. The drive operates at a constant DC-link voltage (e.g., 300 V), with a sampling time of 100 µs and a rated load torque of $T_L =\,$1.2 N·m. All controllers share identical current limits ($\pm$10 A) and a speed reference of 700 rad/s for fair benchmarking.

\begin{figure}
    \centering
    \includegraphics[width=1.0\linewidth]{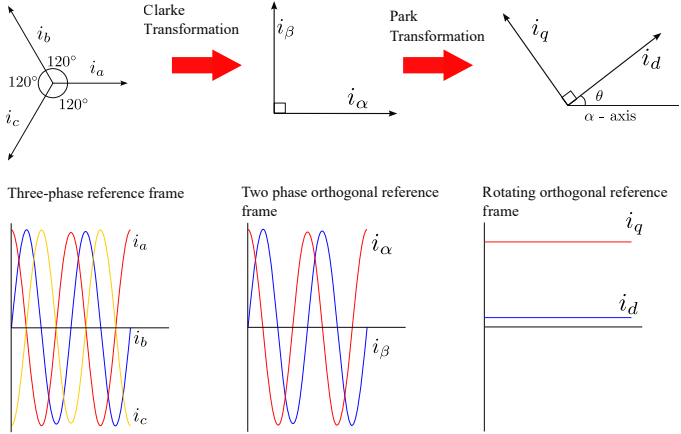}
    \caption{Clarke Park axes}
    \label{fig: transformation}
\end{figure}

This comparative study employs a standard PMSM model as the unified plant for benchmarking. The model is based on a surface-mounted PMSM (SPMSM), where $L_d = L_q$, simplifying the torque dynamics. To ensure a fair comparison, all six SMC variants are tested against this identical plant model. The model is defined in the d-q rotating reference frame. This is achieved by first projecting the three-phase stator quantities $\{a,b,c\}$ onto the stationary $\alpha\beta$ frame (Clarke transform) and then rotating this frame by the electrical angle $\theta_e$ into the rotor-aligned $dq$ frame (Park transform) \cite{omrane4review}, as illustrated in Fig.~\ref{fig: transformation}.  This model operates under standard assumptions, including sinusoidal stator windings and neglecting magnetic saturation. The stator voltage equations of the PMSM are given by:

\begin{align}
\frac{d i_d}{dt} &= \frac{1}{L_d} \left( u_d - R_s i_d + \omega_e L_q i_q \right) \\
\frac{d i_q}{dt} &= \frac{1}{L_q} \left( u_q - R_s i_q - \omega_e (L_d i_d + \psi_f) \right)
\end{align}

The electromagnetic torque $T_e$ produced is defined as
\begin{equation}
T_e = \frac{3}{2} p_n \left[ \psi_f i_q + (L_d - L_q) i_d i_q \right]
\end{equation}

when $L_d = L_q$, as in surface-mounted PMSMs

\begin{equation}
T_e = \frac{3}{2} p_n \psi_f i_q
\end{equation}

Newton's second law governs the mechanical dynamics of the rotor:

\begin{equation}
\label{eq:mechanical}
J \frac{d \omega_r}{dt} = T_e - B \omega_r - T_L
\end{equation}
And, the relationship between electrical and mechanical angular speeds is
\begin{equation}
\omega_e = p_n \omega_r
\end{equation}

Similar modeling is also described in \cite{wang2024disturbance} and \cite{zhang2024improved}.




\begin{table}[t]
\centering
\caption{Nominal parameters of the PMSM used for simulation and benchmarking.}
\label{tab:pmsm_params}
\begin{tabular}{@{}lll@{}}
\toprule
\textbf{Symbol} & \textbf{Description} & \textbf{Value / Unit} \\ 
\midrule
$R_s$    & Stator phase resistance & $0.9~\Omega$ \\
$L_d=L_q$ & $d$–$q$ axis inductance & $8.5~\text{mH}$ \\
$\psi_f$ & Rotor flux linkage & $0.175~\text{Wb}$ \\
$p_n$    & Number of pole pairs & $4$ \\
$J$      & Rotor moment of inertia & $2.8\times10^{-4}~\text{kg·m}^2$ \\
$B$      & Viscous friction coefficient & $1.5\times10^{-4}~\text{N·m·s/rad}$ \\
$T_L$    & Nominal load torque & $1.2~\text{N·m}$ \\
$V_{dc}$ & DC-link voltage & $300~\text{V}$ \\
$f_s$    & Sampling frequency & $10~\text{kHz}$ \\
\bottomrule
\end{tabular}
\end{table}

\section{Sliding Mode Control Design Principles}
\subsection{Design of the Sliding Manifold}
Sliding modes represent distinct evolutions of the controlled system state that occur within a designated subspace of the state space, known as the sliding manifold $(s)$  \cite{FERRARA2026100}. The sliding manifold serves as a subspace in which the system's closed-loop trajectories are enforced \cite{utkin2020road}. A scalar sliding function defines it, 

\begin{equation}
\label{eq:s_manifold}
    s(x) = 0
\end{equation}

\noindent where $x$ is the system's state vector \cite{utkin2020conventional}. This function is chosen so that the system's motion, once constrained to the manifold, is stable and exhibits the desired performance \cite{hossain2017sliding}.

A typical design strategy is the Lyapunov approach, which employs a Lyapunov function. 

\begin{equation}
\label{eq:lyapunov}
   V = \frac{1}{2} s^2 
\end{equation}

\noindent such that $s(x) = 0$ is achieved within a finite time; thus, a typical method is to enforce \cite{yu2017sliding}

\begin{equation*} 
\dot{V} < -\rho\vert s\vert =-\sqrt{2}\rho V^{1/2}, \rho > 0. 
\end{equation*}

\begin{figure}
    \centering
    \includegraphics[width=0.5\linewidth]{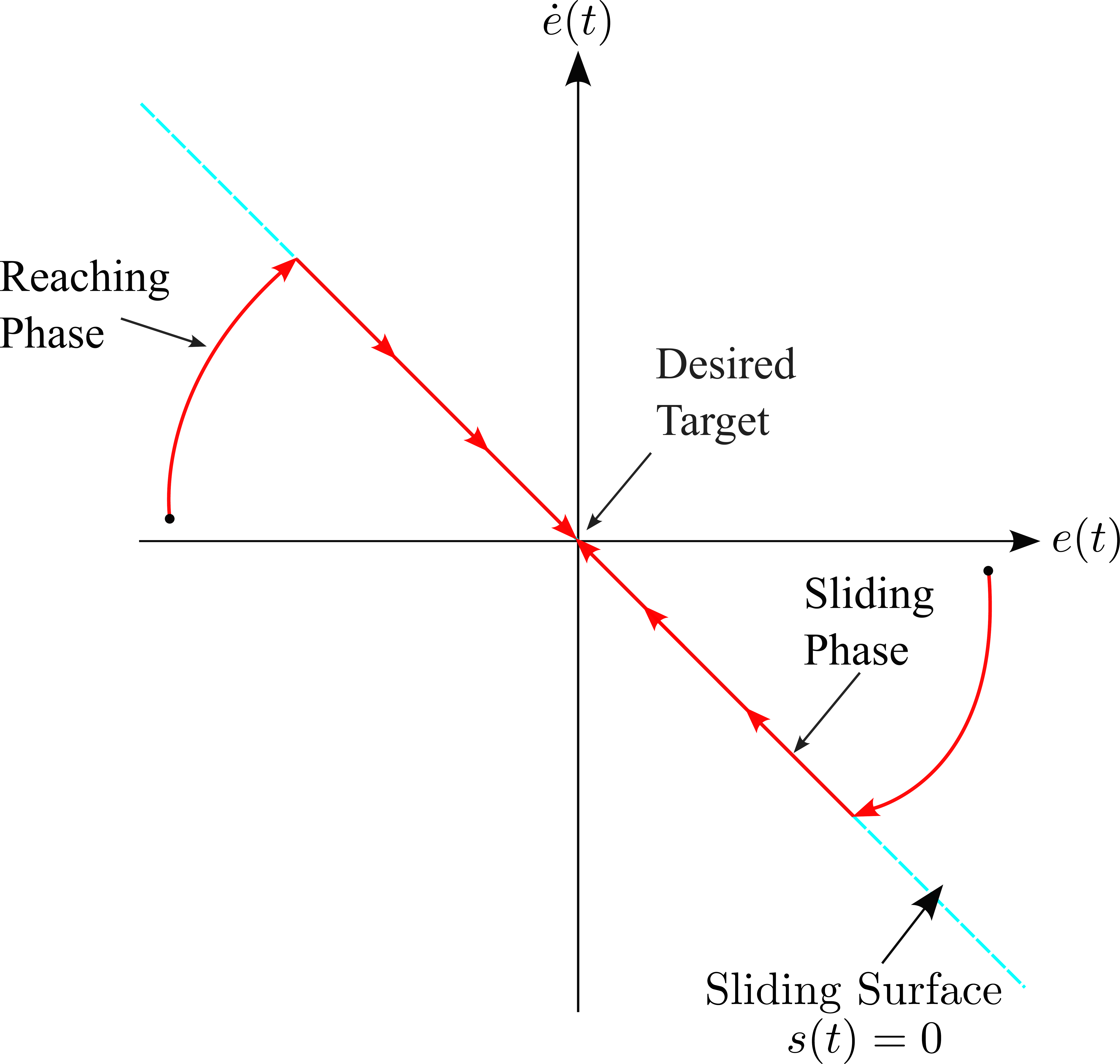}
    \caption{Reaching Phase and Sliding Phase}
    \label{fig:rm_n_sm}
\end{figure}

\subsection{Design of the Control Law}
Subsequently, after designing the manifold, the control law must be developed to enforce the sliding mode \cite{utkin2020road}. The objective is to ensure that any state trajectory starting the manifold \ref{eq:s_manifold} is driven towards it and maintained there, which is known as the reaching mode \cite{mousavi2022sliding} and then transition into the sliding mode \cite{eker2006sliding} as shown in Fig \ref{fig:rm_n_sm}.

To guarantee this, a Lyapunov-based approach is often used, where the control law is designed to make the derivative of a Lyapunov function \ref{eq:lyapunov} negative definite. Such that

\begin{equation}
    s\dot{s} < 0
\end{equation}

To satisfy this condition, a discontinuous control structure is used. Given the general affine nonlinear system:

\begin{equation}
    \dot{x} = f(x,t) + B(x,t) \, u  \;  + \, \zeta(x,t) 
\end{equation}

\noindent such that 
\[
x \in \mathbb{R}^n , \quad u \in \mathbb{R}^m, \quad \zeta(x,t) \in \mathbb{R}^n
\]

where $x \in \mathbb{R}^n$ is the state, $u \in \mathbb{R}^m$ is the control input, and $\zeta(x,t)$ denotes the uncertainties and external disturbances. The discontinuous control law that satisfies the reaching condition is typically defined component-wise as \cite{yu2009sliding}:

\begin{equation}
    u_i =
\begin{dcases}
u_i^{+}(x), & s_i(x) > 0,\\
u_i^{-}(x), & s_i(x) < 0.
\end{dcases}
\end{equation}

Once the system trajectories are on the manifold, the system is in sliding mode \cite{utkin2005sliding}. In this mode, an equivalent control $u_{eq}$ can be induced \cite{utkin2004sliding}. This is the hypothetical continuous control signal that would be required to maintain the state on the manifold \cite{eker2010second}, which means it must satisfy the condition $\dot{s} = 0$.  The time derivative of the sliding variable is:

\begin{equation}
\dot{s} = \frac{\partial s}{\partial x} \dot{x} = \frac{\partial s}{\partial x} \left( f(x,t) + B(x,t)u + \xi(x,t) \right).
\end{equation}

By setting $\dot{s} = 0$, replacing $u$ with $u_{eq}$, and assuming the matrix $(\partial s / \partial x) B(x,t)$ is nonsingular, we can solve for the equivalent control:

\begin{equation}
u_{eq} = - \left( \frac{\partial s}{\partial x} B(x,t) \right)^{-1}\left( \frac{\partial s}{\partial x} \left( f(x,t) + \xi(x,t) \right) \right)
\end{equation}

When the system is in sliding mode, its dynamics are governed by Equation \ref{eq: govern}. If the disturbance $\xi(x,t)$ satisfies the matching condition (i.e., it is in the range space of the control input $B(x,t)$), the motion is then governed by \cite{yu2009sliding}:

\begin{equation}
\label{eq: govern}
\dot{x} =\left[ I - B(x,t)\left( \frac{\partial s}{\partial x} B(x,t) \right)^{-1}\frac{\partial s}{\partial x} \right]f(x,t)
\end{equation}


\subsection{Parameter Selection and Tuning Guidelines}
The control parameters of all SMC variants were tuned according to Lyapunov stability conditions and empirical trade-offs between convergence speed, chattering amplitude, and steady-state accuracy. 
Initial gains were selected to satisfy the reaching condition $s\dot{s} < 0$ under nominal operating conditions, followed by fine adjustments through simulation sweeps. 
The primary tuning considerations for each controller are summarized below.

\begin{itemize}
    \item \textbf{Conventional SMC (CSMC):} Gains $\varepsilon$ and $k$ were chosen to ensure finite-time convergence while avoiding excessive control amplitude. The exponents $a,b\in(0,1)$ were fixed to balance reaching speed and chattering suppression.

    \item \textbf{Integral SMC (ISMC):} The integral weight $\lambda$ was tuned to eliminate steady-state bias caused by constant disturbances. 
    Larger $\lambda$ increases disturbance rejection but may prolong settling time.

    \item \textbf{Terminal SMC (TSMC):} The nonlinear surface parameters $(c,\alpha,k)$ were set to yield finite-time convergence; typically $0.5<\alpha<1$. 
    A small boundary layer $\Delta e$ was applied to approximate the $\operatorname{sgn}(\cdot)$ function and reduce chatter near equilibrium.

    \item \textbf{Fractional-Order SMC (FOSMC):} Fractional orders $(\alpha,\beta)$ were constrained to $0.6\!-\!0.9$ to exploit memory effects without introducing numerical stiffness. 
    The PID-like gains $(k_p,k_i,k_d)$ were tuned to minimize integral indices (ISE, IAE).  

    \item \textbf{Adaptive SMC (ASMC):} The adaptation gains $(\eta_1,\eta_2,\eta_3)$ and limits $(\Omega_r, H)$ were adjusted so that the adaptive switching gain $\Omega(t)$ increased rapidly during significant errors but decayed near equilibrium, minimizing chattering. 

    \item \textbf{Super-Twisting SMC (STSMC):} The gains $(k_1,k_2)$ were selected to satisfy the finite-time condition $k_1^2 > 4k_2$, ensuring convergence under bounded disturbances. 
    Parameter $c$ shaped the surface slope and controlled the transient response. 
    Both gains were fine-tuned empirically to achieve smooth control effort without overshoot.
\end{itemize}

Overall, all controllers were tuned to ensure stability margins under nominal parameters while maintaining comparable rise and settling times, enabling a fair quantitative comparison across all six SMC strategies.

\begin{table}[t]
\centering
\caption{Controller tuning parameters used in the simulation benchmark.}
\label{tab:ctrl_tuning_values}
\resizebox{\linewidth}{!}{%
\begin{tabular}{@{}lcccccc@{}}
\toprule
\textbf{Controller} & \textbf{Key Gains / Parameters} & \textbf{Values (Nominal)} \\ 
\midrule
CSMC  & $\varepsilon,\,k,\,a,\,b$ & $15,\,5,\,0.6,\,0.8$ \\[3pt]
ISMC  & $\lambda,\,\varepsilon,\,k,\,a,\,b$ & $30,\,15,\,5,\,0.6,\,0.8$ \\[3pt]
TSMC  & $c,\,\alpha,\,k,\,\Delta e$ & $20,\,0.7,\,10,\,0.02$ \\[3pt]
FOSMC & $k_p,\,k_i,\,k_d,\,\alpha,\,\beta,\,w,\,k_s$ & $1.0,\,30,\,0.002,\,0.8,\,0.7,\,40,\,8$ \\[3pt]
ASMC  & $c,\,\eta_1,\,\eta_2,\,\eta_3,\,\Omega_r,\,H$ & $15,\,2.0,\,1.5,\,0.2,\,1.0,\,50$ \\[3pt]
STSMC & $c,\,k_1,\,k_2$ & $15,\,8,\,3$ \\ 
\bottomrule
\end{tabular}}
\end{table}

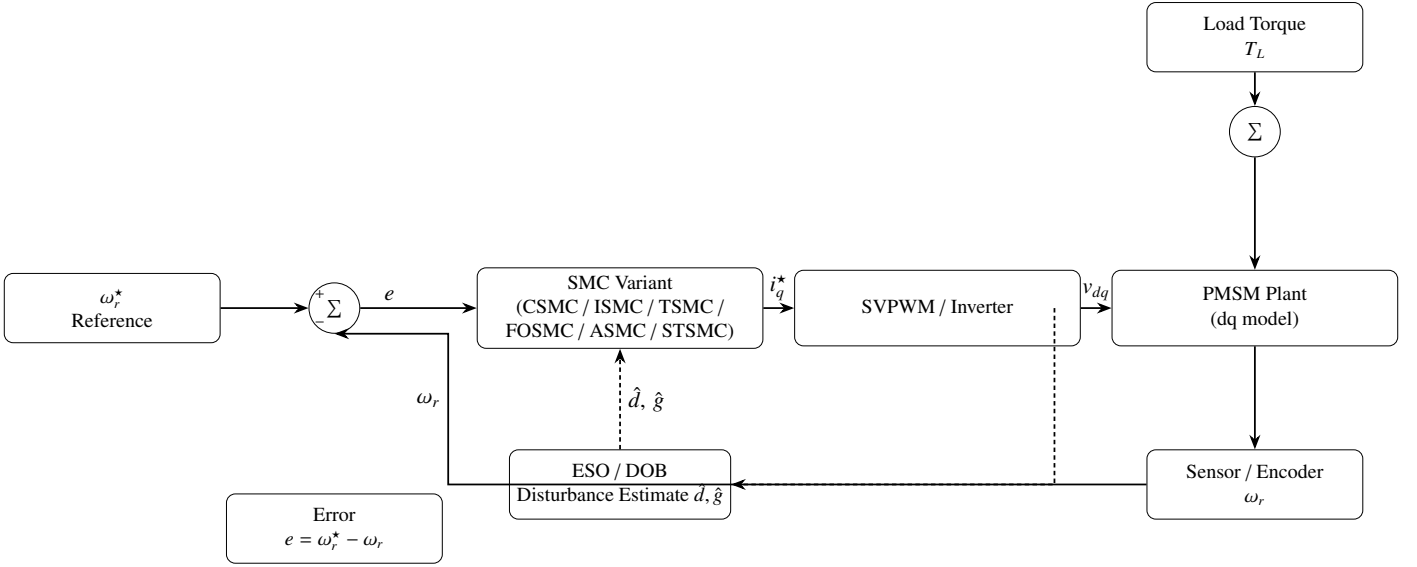
\begin{figure*}[t]
\resizebox{\textwidth}{!}{%
\centering
\begin{tikzpicture}[x=1cm,y=1cm]
\tikzset{
  >={Stealth[length=2mm,width=1.6mm]},
  arrow/.style={-Stealth,thick},
  dashedarrow/.style={dash pattern=on 2pt off 1.5pt,-Stealth,thick},
  block/.style={draw,rounded corners,minimum height=1.2cm,minimum width=4.5cm,align=center,font=\small},
  smallblock/.style={draw,rounded corners,minimum height=1.1cm,minimum width=3.4cm,align=center,font=\small},
  sum/.style={draw,circle,minimum size=8mm,inner sep=0pt,font=\small}
}

\def\Ydist{2.8}

\node[smallblock] (ref) at (-8,0) {$\omega_r^\star$\\Reference};
\node[sum]        (sum1) at (-4.5,0) {$\sum$};
\node[block]      (ctrl) at (0,0) {SMC Variant\\(CSMC / ISMC / TSMC /\\FOSMC / ASMC / STSMC)};
\node[block]      (inv)  at (5,0) {SVPWM / Inverter};
\node[block]      (plant)at (10,0) {PMSM Plant\\(dq model)};

\node[sum]        (sumd) at (10,\Ydist) {$\sum$};
\node[smallblock] (dist) at (10,\Ydist+1.5) {Load Torque\\$T_L$};

\node[smallblock] (eso)  at (0,-\Ydist) {ESO / DOB\\Disturbance Estimate $\hat d, \hat g$};
\node[smallblock] (meas) at (10,-\Ydist) {Sensor / Encoder\\$\omega_r$};
\node[smallblock] (err)  at (-4.5,-\Ydist-0.7) {Error\\$e = \omega_r^\star - \omega_r$};

\draw[arrow] (ref.east) -- (sum1.west);
\draw[arrow] (sum1.east) -- node[above,near start] {$e$} (ctrl.west);
\draw[arrow] (ctrl.east) -- node[above] {$i_q^\star$} (inv.west);
\draw[arrow] (inv.east) -- node[above] {$v_{dq}$} (plant.west);

\draw[arrow] (dist.south) -- (sumd.north);
\draw[arrow] (sumd.south) -- (plant.north);

\draw[arrow] (plant.south) -- (meas.north);
\draw[arrow] (meas.west) -- ++(-11,0) |- (sum1.south)
           node[pos=0.28,left] {$\omega_r$};

\draw[dashedarrow] (plant.west)++(-0.9,0) |- (eso.east);
\draw[dashedarrow] (eso.north) -- (ctrl.south)
           node[midway,right] {$\hat d,\,\hat g$};

\node at ($(sum1.center)+(-2.3mm, 2.2mm)$) {\scriptsize $+$};
\node at ($(sum1.center)+(-2.3mm,-2.2mm)$) {\scriptsize $-$};

\end{tikzpicture}}
\caption{Closed-loop PMSM speed control system with SMC variant, ESO/DOB compensation, inverter, PMSM plant, load torque disturbance, and explicit feedback path. Distances are increased to maintain clarity and prevent overlaps.}
\label{fig:block_pmsm_smc}
\end{figure*}

\section{Simulation}
To quantitatively evaluate the performance of various SMC techniques applied to PMSMs, a series of high-fidelity numerical simulations was conducted using a standardized plant model and simulation environment based on the setup in \cite{wang2019new}. Each controller was tested under identical conditions to ensure a fair and objective comparison. All simulations were implemented in MATLAB/Simulink using a fixed-step Runge–Kutta solver (step size = 10 µs).

The simulation study covers CSMC, ISMC, STSMC, TSMC, ASMC, and FOSMC. The system model incorporates feedforward compensation via an extended state observer (ESO) in all cases, capturing the dynamic effects of lumped disturbances such as load torque and model uncertainties. Performance assessment was carried out under two distinct scenarios: (i) nominal operation without external disturbances and (ii) operation under torque disturbances applied at specific time instances. Evaluation criteria included rotor speed tracking accuracy, convergence time, robustness against disturbances, and chattering mitigation. Classical time-domain specifications (overshoot, rise time, steady-state error) and integral error indices (ISE, IAE, ITSE, ITAE) are also computed to support a rigorous comparison of each SMC strategy.

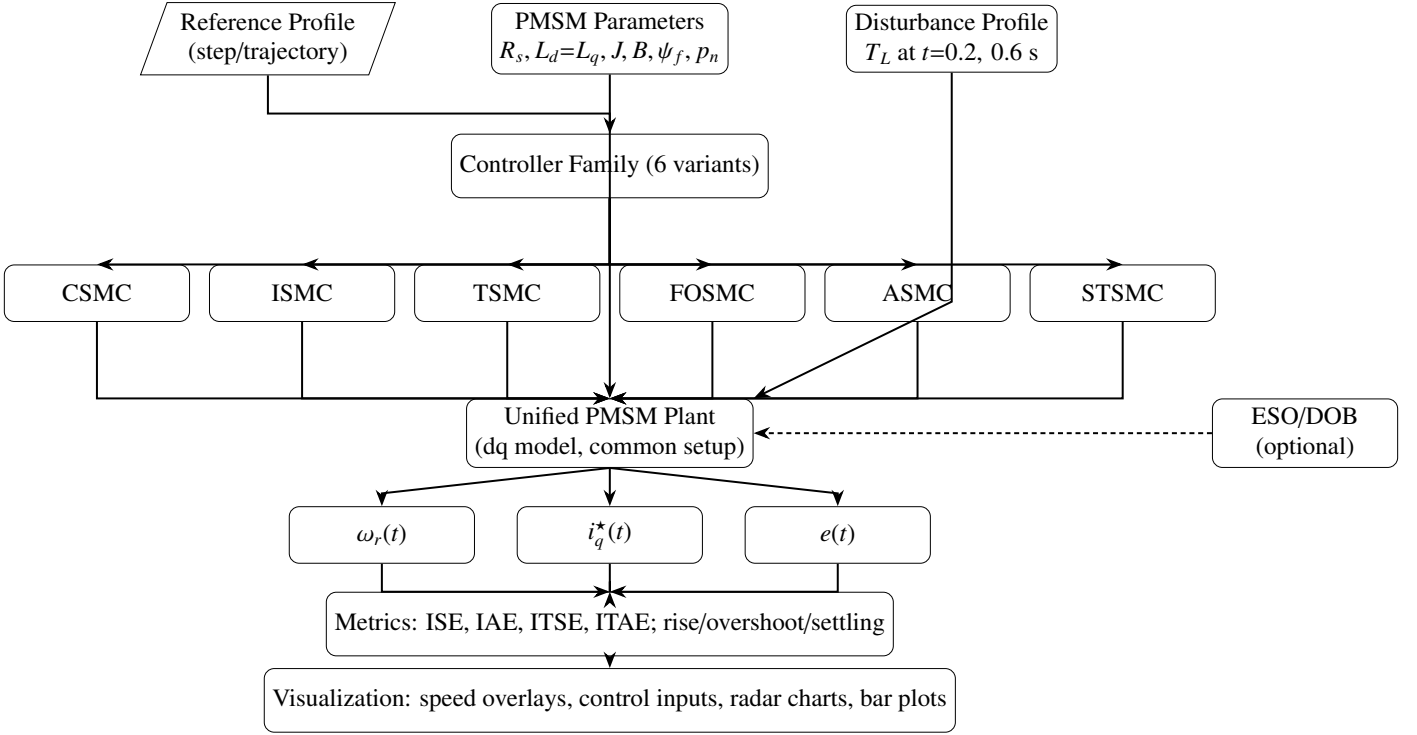
\begin{figure*}[t]
\resizebox{\textwidth}{!}{%
\centering
\begin{tikzpicture}[x=0.8cm,y=0.9cm]

\node[io]        (ref)   at (-6.0, 4.0) {Reference Profile\\(step/trajectory)};
\node[smallblock](params)at ( 0.0, 4.0) {PMSM Parameters\\$R_s, L_d{=}L_q, J, B, \psi_f, p_n$};
\node[smallblock](dist)  at ( 6.0, 4.0) {Disturbance Profile\\$T_L$ at $t{=}0.2,\,0.6$ s};

\node[block]     (ctrlfam) at (0.0, 2.0) {Controller Family (6 variants)};

\node[smallblock](csmc)  at (-9.0, 0.0) {CSMC};
\node[smallblock](ismc)  at (-5.4, 0.0) {ISMC};
\node[smallblock](tsmc)  at (-1.8, 0.0) {TSMC};
\node[smallblock](fosmc) at ( 1.8, 0.0)  {FOSMC};
\node[smallblock](asmc)  at ( 5.4, 0.0)  {ASMC};
\node[smallblock](stsmc) at ( 9.0, 0.0)  {STSMC};

\node[smallblock](eso)   at (12.2, -2.2) {ESO/DOB\\(optional)};

\node[block]     (plant) at (0.0, -2.2) {Unified PMSM Plant\\(dq model, common setup)};

\node[smallblock](speed) at (-4.0, -3.8) {$\omega_r(t)$};
\node[smallblock](iq)    at ( 0.0, -3.8) {$i_q^\star(t)$};
\node[smallblock](err)   at ( 4.0, -3.8) {$e(t)$};

\node[block]     (metrics) at (0.0, -5.2) {Metrics: ISE, IAE, ITSE, ITAE; rise/overshoot/settling};
\node[block]     (plots)   at (0.0, -6.4) {Visualization: speed overlays, control inputs, radar charts, bar plots};

\draw[arrow] (ref.south) -- ++(0,-0.6) -| (ctrlfam.north);
\draw[arrow] (params.south) -- ++(0,-3.6) -- (plant.north);
\draw[arrow] (dist.south) -- ++(0,-3.6) -- (plant.north east);

\foreach \n in {csmc,ismc,tsmc,fosmc,asmc,stsmc}{
  \draw[arrow] (ctrlfam.south) -- ++(0,-0.6) |- (\n.north);
}

\foreach \n in {csmc,ismc,tsmc,fosmc,asmc,stsmc}{
  \draw[arrow] (\n.south) -- ++(0,-0.6) |- (plant.north);
}

\draw[dashedarrow] (eso.west) -- (plant.east);

\draw[arrow] (plant.south) -- ++(-4, -0.4) -- (speed.north);
\draw[arrow] (plant.south) -- (iq.north);
\draw[arrow] (plant.south) -- ++( 4, -0.4) -- (err.north);

\foreach \n in {speed,iq,err}{
  \draw[arrow] (\n.south) -- ++(0,-0.4) |- (metrics.north);
}
\draw[arrow] (metrics.south) -- (plots.north);

\end{tikzpicture}}
\caption{Compact unified benchmarking pipeline: identical plant, parameters, and disturbance applied to six SMC variants, with shared metrics and Visualization.}
\label{fig:benchmark_flow}
\end{figure*}

\subsection{Conventional Sliding-Mode Control}\label{sec:Trad_sim_smc}

The baseline simulation evaluates a CSMC strategy for PMSM speed regulation. 
The plant model and parameters are based on \cite{wang2019new}. The error is defined as 
\begin{equation}
\label{eq:error}
   e \;=\; \omega_r^\star - \omega_r ,
\end{equation}
where $\omega_r^\star$ and $\omega_r$ denote the reference and measured rotor speeds. While the sliding surface is defined as 
\begin{equation}
\label{eq:c_surface}
s = \dot{e} + c\,e,
\end{equation}

Let the PMSM speed dynamics be written as
\begin{equation}
    \dot{\omega}_r \;=\; \chi\, i_q \;+\; g \;+\; d, 
    \qquad \chi := \frac{3p\psi_f}{2J},
\end{equation}
with $i_q$ the torque-producing current, $p$ the pole-pairs, $\psi_f$ the flux linkage, $J$ the rotor inertia, and $g,d$ the matched uncertainty and lumped disturbance (with estimates $\hat g,\hat d$).

Keeping the reaching law as in the base paper, we use
\begin{equation}
    \dot{s} \;=\; 
    -\,\varepsilon\,|x|^{a}\,\operatorname{sgn}(s)\;
    -\,k\,|s|^{\,b\,\operatorname{sgn}(|s|-1)}\,s,
\end{equation}

where

\begin{center}
   $ \varepsilon>0,\; k>0,\; 0<a<1,\; 0<b<1,$
\end{center}
and in our setting take $x=e$ (i.e., $x$ equals the speed error).

Since $\dot s = \dot e = \dot{\omega}_r^\star - \dot{\omega}_r = \dot{\omega}_r^\star - \chi i_q - g - d$, equating with the reaching law and solving for $i_q$ gives the torque–current command
\begin{equation}
    i_q^{\star}(t)
    \;=\;
    \frac{1}{\chi}\!\left[
        \dot{\omega}_r^\star
        - \hat g
        - \hat d
        + \varepsilon\,|x|^{a}\operatorname{sgn}(s)
        + k\,|s|^{\,b\,\operatorname{sgn}(|s|-1)}\,s
    \right],
\end{equation}
which preserves the finite-time reaching and chattering-attenuation properties reported in the base paper while using the simpler surface $s=e$.

\begin{figure*}[htbp]
  \centering
  \hspace*{-0.18\textwidth}
  \begin{subfigure}[b]{0.8\textwidth}      
    \centering
    \includegraphics[ width=\linewidth,trim=12bp 8.5cm 12bp 8.5cm, clip]{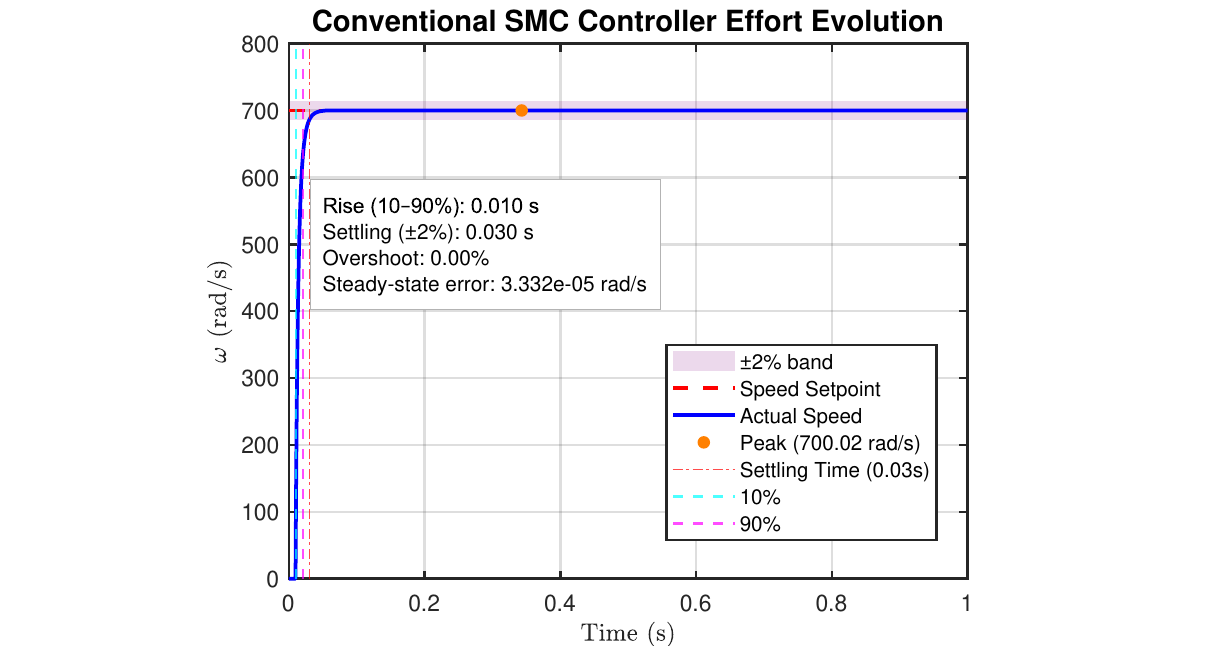}
    \caption{Without external disturbance.}
    \label{fig:trad_nodist}
  \end{subfigure}%
  \hspace*{-0.25\textwidth}                 
  \begin{subfigure}[b]{0.8\textwidth}
    \centering
    \includegraphics[ width=\linewidth, trim=12bp 8.5cm 12bp 8.5cm, clip]{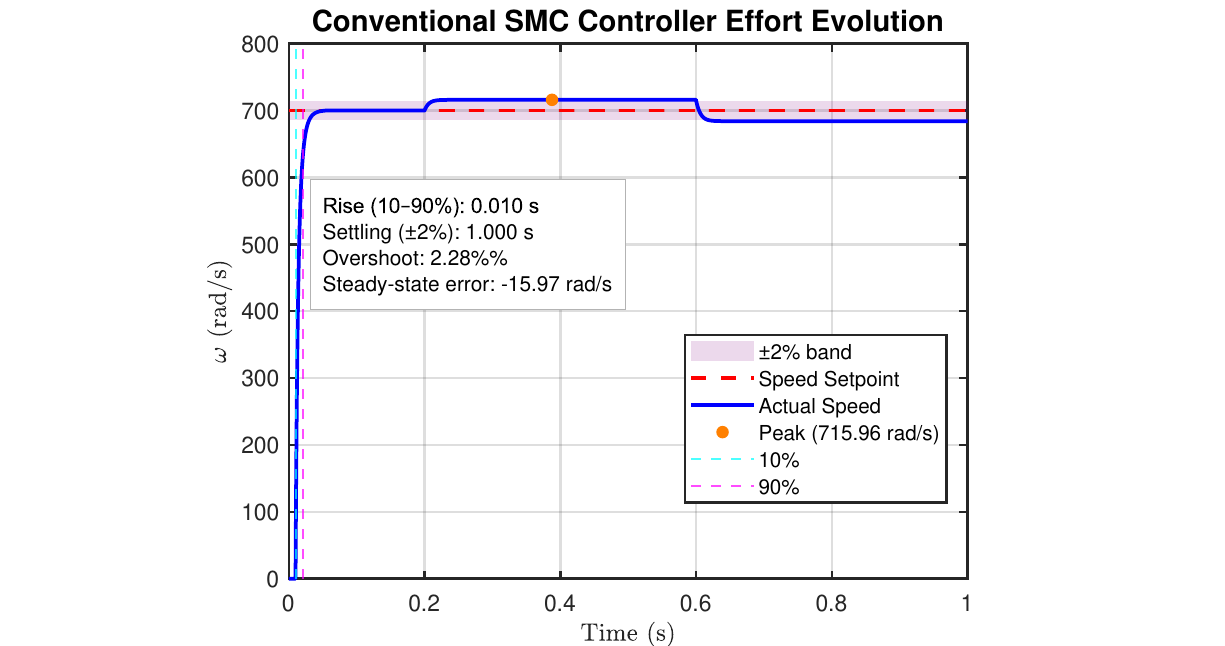}
    \caption{With disturbance at \(t = 0.2\,\text{s}\) and \(0.6\,\text{s}\).}
    \label{fig:trad_dist}
  \end{subfigure}
  \caption{Rotor-speed response under Conventional SMC..}
  \label{fig:trad_pair}
\end{figure*}

Figure~\ref{fig:trad_pair} illustrates the rotor-speed response under conventional SMC. Two cases are presented: (i) nominal operation without external disturbance and (ii) disturbance-injected conditions with abrupt torque variations. 
Under a single step command, the conventional SMC achieves fast tracking without disturbance, with a rise time of \(0.01025\,\text{s}\), a settling time (2\% band) of \(0.02035\,\text{s}\), a negligible overshoot of \(0.003\%\), and \(\approx 0\) steady-state error. In the disturbed case, the rise time remains unchanged \((0.01025\,\text{s})\) but the response develops a persistent bias, yielding an overshoot of \(2.28\%\) and a steady-state error of \(+15.97\) relative to the \(700\) setpoint (\(\approx 2.28\%\)). Consequently, the output does not re-enter the \(\pm 2\%\) band within the simulation window, indicating that while reaching dynamics are rapid, disturbance rejection and steady-state accuracy would benefit from integral/terminal action or boundary-layer tuning to remove the bias and restore band-limited regulation.

Figures \ref{fig:trad_nodist} and \ref{fig:trad_dist} show the speed response of the conventional SMC applied to PMSM speed control. In the absence of disturbances, the controller achieves rapid convergence to the setpoint with a smooth steady-state profile. Under torque disturbances, it maintains regulation but exhibits increased speed ripple, reflecting the influence of the discontinuous switching action on disturbance rejection.

\subsection{Integral Sliding Mode Control}

In this simulation, the ISMC strategy is applied to the same PMSM model and parameter set used in \cite{wang2019new}. ISMC is designed to eliminate the reaching phase by initiating system trajectories directly on the sliding manifold, thereby maintaining robustness from the initial time.

Let the speed-tracking error be defined as \(e = \omega_r^{\star} - \omega_r\), where \(\omega_r^{\star}\) is the reference speed. The integral sliding surface is constructed as:
\[
s(t) = e(t) + \lambda \int_{0}^{t} e(\tau)\,d\tau,
\]
where \(\lambda > 0\) determines the contribution of the integral term. Differentiating the surface yields:
\[
\dot{s} = \dot{e} + \lambda e.
\]

To ensure fair comparison with conventional SMC, the same nonlinear reaching law is retained:
\[
\dot{s} = -\varepsilon |e|^{a}\operatorname{sgn}(s)
          - k\,|s|^{b\operatorname{sgn}(|s|-1)}\,s,
\]
where \(\varepsilon, k > 0\) and \(0 < a, b < 1\) are design parameters governing convergence speed and chattering suppression.

The corresponding control law for the \(q\)-axis current is defined as:
\[
i_q^{\star} =
\frac{1}{\chi} \left(
    \dot{\omega}_r^{\star}
    + \varepsilon |e|^a \operatorname{sgn}(s)
    + k |s|^{b \operatorname{sgn}(|s| - 1)} s
    + \lambda e
\right),
\]
where \(\chi = \frac{3p\psi_f}{2J}\) represents the electromechanical coupling factor.

\begin{figure*}[hbtp]
  \centering
  \hspace*{-0.18\textwidth}
  \begin{subfigure}[b]{0.8\textwidth}      
    \centering
    \includegraphics[ width=\linewidth, trim=12bp 8.5cm 12bp 8.5cm, clip]{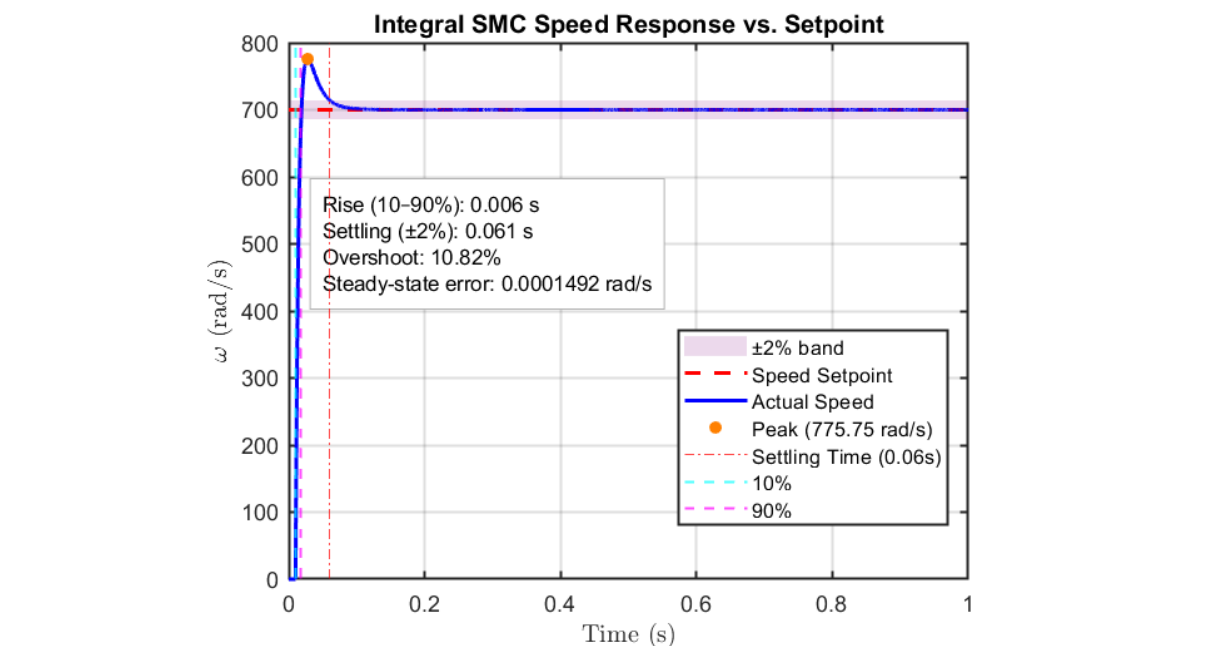}
    \caption{No external disturbance  Integral.}
    \label{fig:ismc_nodist}
  \end{subfigure}%
  \hspace*{-0.25\textwidth}                 
  \begin{subfigure}[b]{0.8\textwidth}
    \centering
    \includegraphics[ width=\linewidth, trim=12bp 8.5cm 12bp 8.5cm, clip]{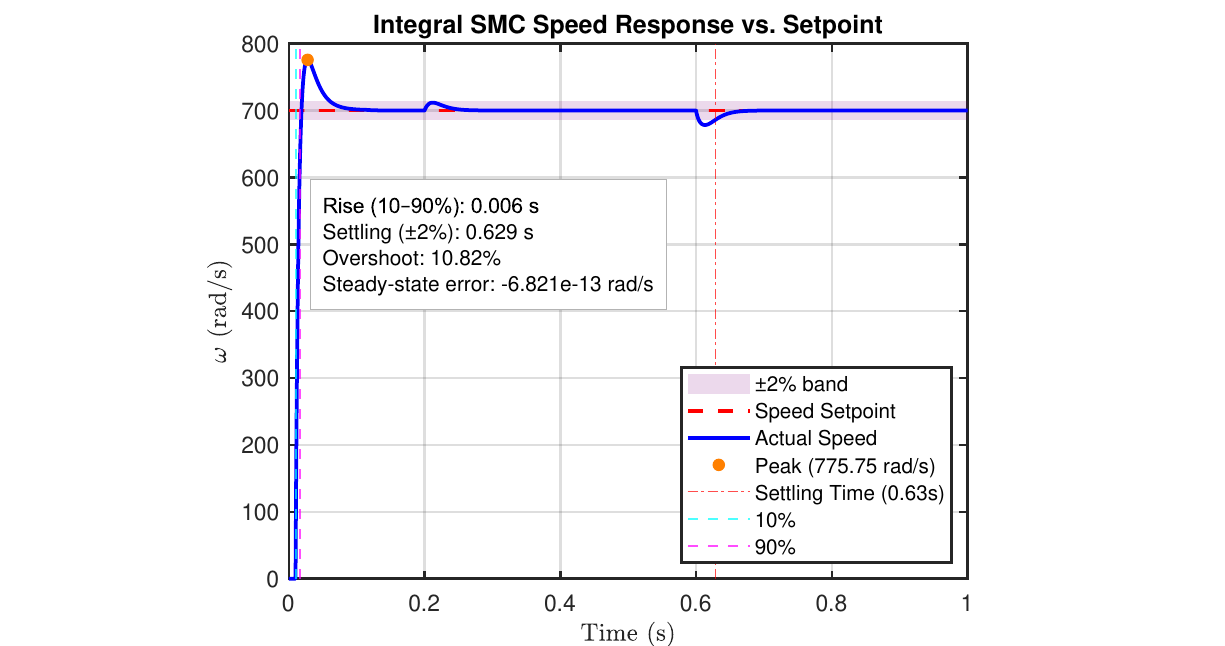}
   \caption{External disturbance applied at \(t=0.2\) s and \(t=0.6\) s..}
    \label{fig:ismc_dist}
  \end{subfigure}
  \caption{Rotor-speed response under integral SMC..}
  \label{fig:ismc_pair}
  \end{figure*}

Figure~\ref{fig:ismc_pair} presents the rotor speed response under ISMC, evaluated both in the absence and presence of external torque disturbances. The integral action helps reject initial modeling errors, and the nonlinear reaching law improves transient robustness while mitigating chattering.

Figures \ref{fig:ismc_nodist} and \ref{fig:ismc_dist} illustrate the PMSM speed response under Integral SMC. The integral sliding mode controller performance was evaluated with one setpoint command in the presence and absence of disturbance.
Under a single-step command, the ISMC achieves accurate tracking with zero steady-state error in both cases. Without disturbance, the speed response exhibits a rise time of \(0.00621\,\text{s}\), a settling time (2\% band) of \(0.05066\,\text{s}\), and an overshoot of \(10.82\%\). With disturbance, the rise time and overshoot remain essentially unchanged \((0.00621\,\text{s},\,10.82\%)\), but the settling time enlarges to \(0.62865\,\text{s}\) (about \(12.4\times\) slower). Overall, ISMC preserves steady-state accuracy, while disturbances primarily tax recovery time. Tuning the boundary layer/reaching law (e.g., gain schedules or a thicker boundary layer) could shorten settling time without sacrificing robustness.

\subsection{Terminal Sliding-Mode Control}
Terminal Sliding-Mode Control (TSMC) introduces finite-time convergence properties by leveraging a nonlinear sliding surface that accelerates error reduction as the state approaches equilibrium. In this simulation, the TSMC strategy is applied to the PMSM plant, using the parameters described in \cite{wang2019new}.

Let the speed-tracking error be defined as \(e = \omega_r^{\star} - \omega_r\). Following the design proposed in \cite{tian2024integrated}, the sliding surface and control law are constructed as follows:
\begin{equation}
\begin{aligned}
  s &= \dot{e} + c\,|e|^{\alpha} \operatorname{sat}(e), \qquad 0 < \alpha < 1, \\[4pt]
  \dot{s} &= -k\,\operatorname{sgn}(s), \qquad k > 0, \\[4pt]
  i_q^{\star} &= \frac{1}{\chi} \left(
    c\,|e|^{\alpha} \operatorname{sat}(e)
    + k \int \operatorname{sgn}(s)\,dt
  \right), \qquad
  \chi = \frac{3p\psi_f}{2J},
\end{aligned}
\end{equation}
where \(c\), \(\alpha\), and \(k\) are tuning parameters, and \(\chi\) encapsulates the electromechanical coupling constant.

To reduce chattering near the origin, the discontinuous sign function is approximated using a continuous saturation function:
\begin{equation}
  \operatorname{sat}(e) =
  \begin{cases}
    1, & e > \Delta_e, \\[4pt]
    e, & |e| \le \Delta_e, \\[4pt]
    -1, & e < -\Delta_e,
  \end{cases}
  \qquad \Delta_e > 0.
\end{equation}

\begin{figure*}[htbp]
      \hspace*{-0.18\textwidth}
  \begin{subfigure}[b]{0.8\textwidth}
    \centering
   \includegraphics[width=\linewidth, trim=12bp 8.5cm 12bp 8.75cm, clip]{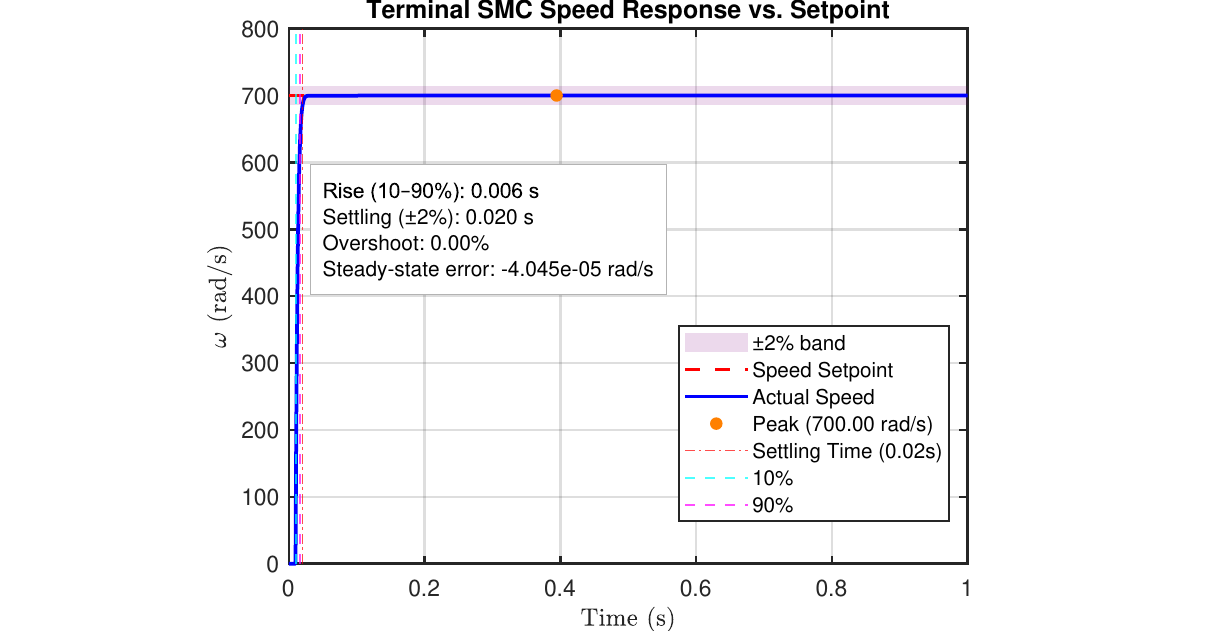}
    \caption{No external disturbance.}
    \label{fig:tsmc_nodist}
  \end{subfigure}%
  \hspace*{-0.25\textwidth}                 
  \begin{subfigure}[b]{0.8\textwidth}
    \centering
    \includegraphics[ width=\linewidth, trim=12bp 8.4cm 12bp 8.4cm, clip]
            {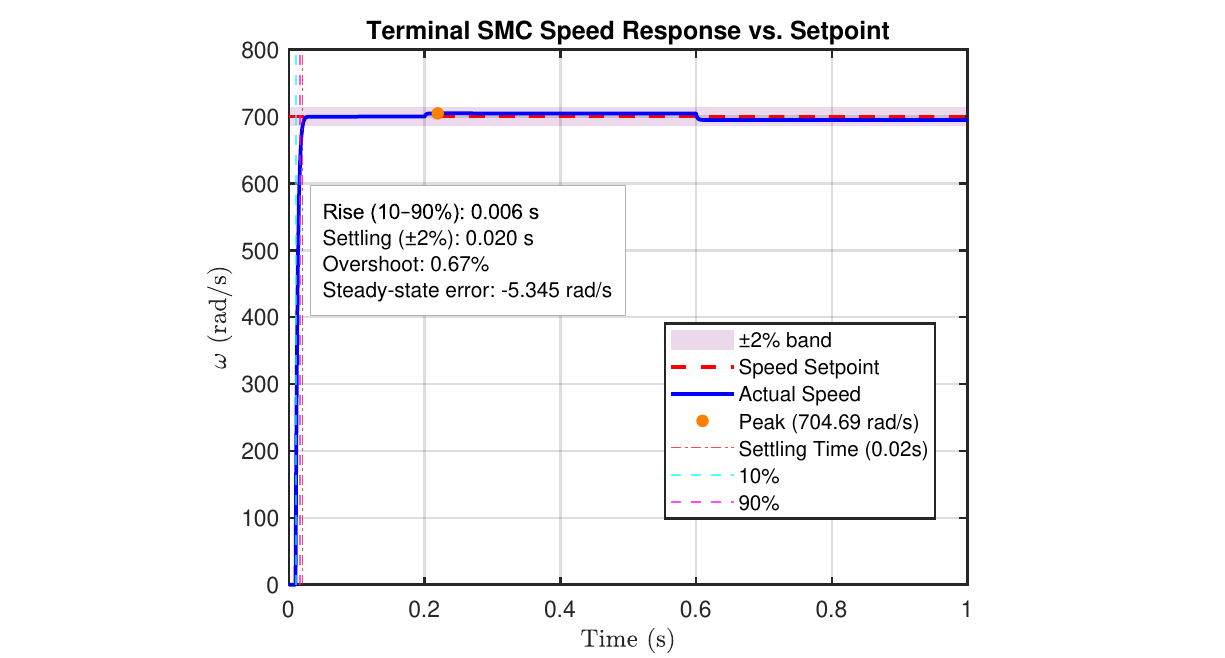}
   \caption{External disturbance applied at \(t=0.2\) s and \(t=0.6\) s..}
    \label{fig:tsmc_dist}
  \end{subfigure}
   \caption{Rotor-speed response under terminal SMC..}
  \label{fig:tsmc_pair}
\end{figure*}

The terminal sliding surface enhances convergence speed for significant errors while maintaining smooth control near equilibrium, thanks to the use of a saturation boundary layer. Figure~\ref{fig:tsmc_pair} shows the rotor-speed responses both without disturbance and under abrupt load changes.

Figures \ref{fig:tsmc_nodist} and \ref{fig:tsmc_dist} depict the speed response of the terminal sliding mode controller.   
For a single-step command, the terminal SMC achieves fast, accurate tracking with zero steady-state error in both cases. Without disturbance, the response shows a rise time of \(0.00628\,\text{s}\), settling time (2\% band) of \(0.01047\,\text{s}\), and negligible overshoot (\(0.0005\%\)). With disturbance, rise and settling remain comparable \((0.00612\,\text{s},\,0.00975\,\text{s})\) with a modest overshoot of \(1.44\%\). Importantly, at the disturbance instants the response stays within the \(\pm 2\%\) band around the \(700\) setpoint at all times; the largest local excursions are a small overshoot of \(+4.70\) (\(\approx 0.67\%\)) near \(t=0.219\,\text{s}\) and a small undershoot of \(-5.19\) (\(\approx 0.74\%\)) near \(t=0.650\,\text{s}\). Thus, the disturbance primarily perturbs the peak value while the terminal dynamics preserve rapid settling and band-limited regulation.

\subsection{Adaptive Sliding-Mode Control (ASMC)}

To enhance convergence performance and minimize chattering, an ASMC strategy is employed in this simulation. The PMSM model, plant parameters, and sampling configuration remain identical to those used in the previous section. 

Define the speed-tracking error as\(e = \omega_r^{\star} - \omega_r\), and construct the sliding surface as in \ref{eq:c_surface}:

where \(c > 0\) is a design constant governing convergence speed.

The adaptive mechanism is based on the scheme proposed in \cite{junejo2020adaptive}, wherein the switching gain \(\Omega(t)\) is adjusted in real time based on the instantaneous values of \(s\) and \(\dot{s}\). The reaching law is given by:
\begin{equation}
\dot{s} = -\Omega\,\operatorname{sgn}\left(s + \frac{\dot{s}}{2\Omega_r|\dot{s}|}\right),
\end{equation}
where the gain \(\Omega\) evolves according to:
\begin{equation}
\Omega =
\begin{cases}
\eta_1 |s| - \eta_2 |\dot{s}| + \eta_3, & \Omega > H, \\[2pt]
M_{\min} + \eta_3, & \Omega < \Omega_r, \\[2pt]
0, & \text{otherwise},
\end{cases}
\end{equation}
with \(\eta_1, \eta_2, \eta_3 > 0\) as adaptation rate constants. The threshold \(\Omega_r\) sets a lower bound on the switching gain, while \(H\) defines an upper limit. This formulation increases the gain when the system is far from the sliding surface, accelerating convergence, and reduces it near the origin to suppress chattering.

The corresponding control law for the torque-generating current \(i_q^{\star}(t)\) is expressed as:
\begin{equation*}
i_q^{\star}(t) =
\frac{1}{\chi} \left(
    \dot{\omega}_r^{\star}
    + c\,e
    + \int \Omega\,\operatorname{sgn}\left(s + \frac{\dot{s}}{2\Omega_r|\dot{s}|}\right)\,dt
\right),
\quad \chi = \frac{3p\psi_f}{2J}
\end{equation*}

This adaptive structure dynamically balances aggressive convergence when the tracking error is significant with smooth, low-gain control near the steady state, thus improving disturbance rejection while suppressing high-frequency oscillations.

\begin{figure*}[htbp]
     \hspace*{-0.18\textwidth}
  \begin{subfigure}[b]{0.8\textwidth}     
    \centering
    \includegraphics[ width=\linewidth, trim=12bp 8.5cm 12bp 8.5cm, clip]{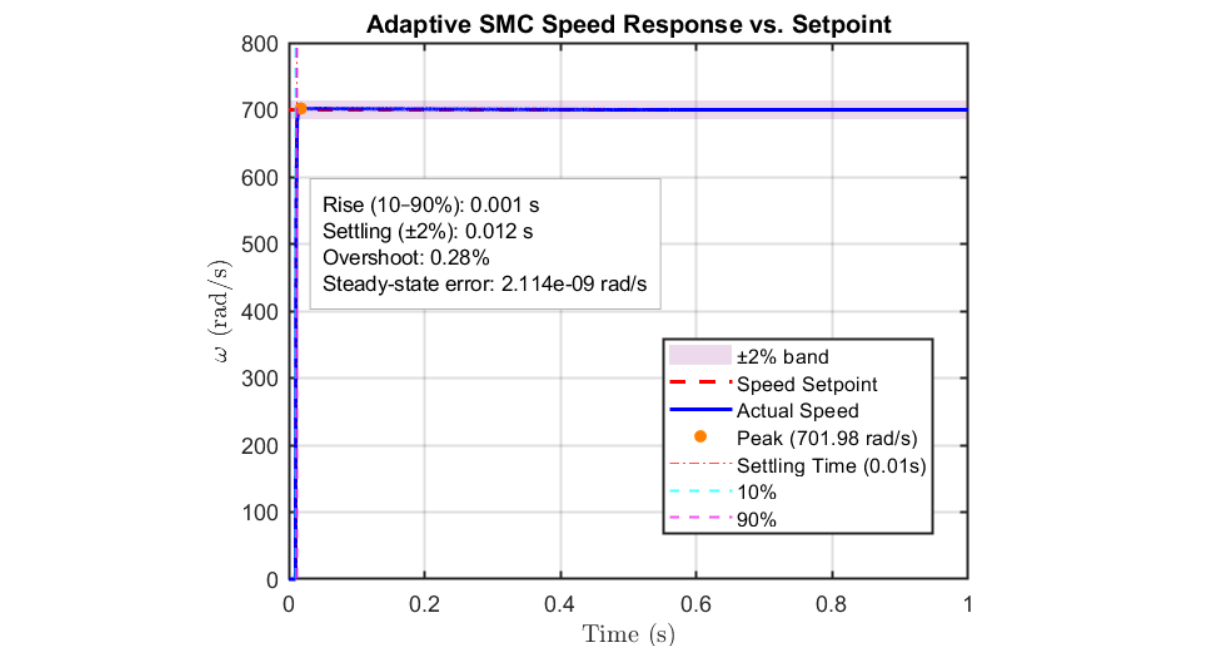}
    \caption{No external disturbance.}
    \label{fig:asmc_nodist}
  \end{subfigure}%
  \hspace*{-0.25\textwidth}                 
  \begin{subfigure}[b]{0.8\textwidth}
    \centering
   \includegraphics[ width=\linewidth, trim=12bp 8.5cm 12bp 8.5cm, clip]{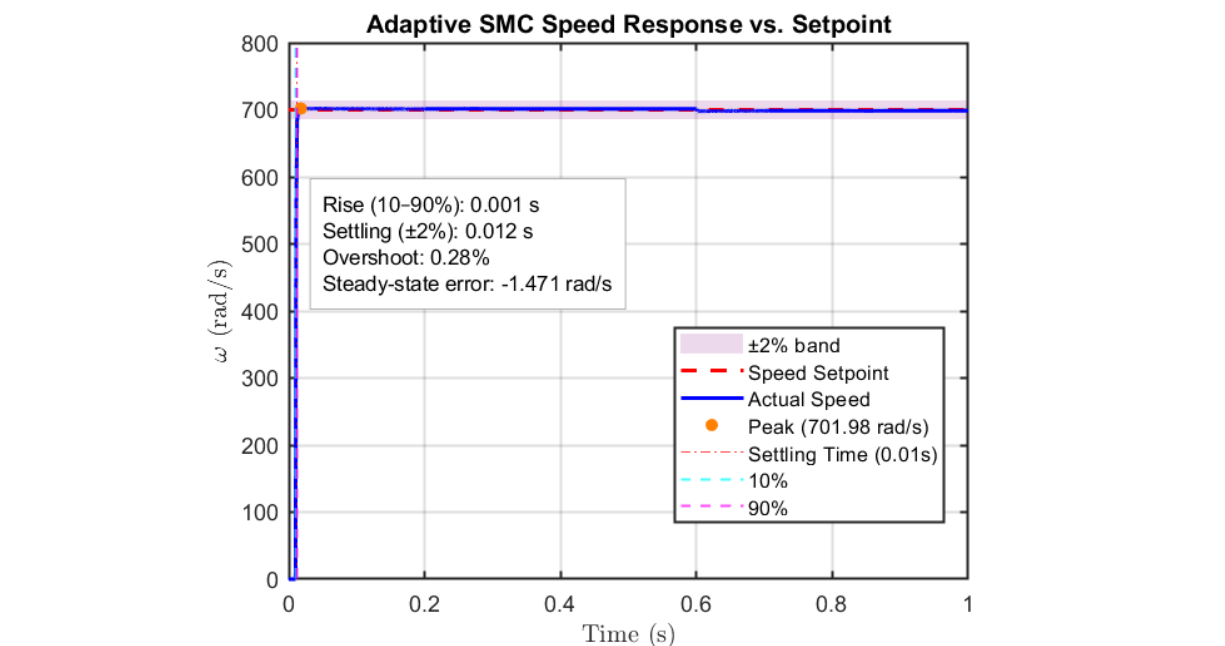}
    \caption{External disturbance applied at \(t=0.2\) s and \(t=0.6\) s..}
    \label{fig:asmc_dist}
  \end{subfigure}
  \caption{Rotor-speed response under adaptive SMC.}
  \label{fig:asmc_pair}
\end{figure*}

Figures \ref{fig:asmc_nodist} and \ref{fig:asmc_dist} depict the speed response of the adaptive sliding mode controller.
Under a single step command, the adaptive controller exhibits high-speed transients with near-zero steady-state error in the disturbance-free case, achieving a rise time of \(0.00133\,\text{s}\), a settling time (2\% band) of \(0.00237\,\text{s}\), and an overshoot of \(0.283\%\). With external disturbances, the global rise and settling remain essentially unchanged \((0.00133\,\text{s},\,0.00237\,\text{s})\), while the final offset is slight \((e_{ss}\approx 1.47)\). Local inspection around disturbance instants shows only band-limited peaks: at \(t=0.2\,\text{s}\) a mild overshoot of \(+1.52\) (\(\approx 0.22\%\) of a \(700\) setpoint), and at \(t=0.6\,\text{s}\) a small undershoot of \(-2.19\) (\(\approx 0.31\%\)), both well within the \(\pm 2\%\) tolerance. Overall, the adaptive scheme preserves millisecond-scale settling and tight regulation in the presence of disturbances, with only minor peak excursions and a negligible steady-state bias under load.

\subsection{Fractional-Order Sliding-Mode Control (FOSMC)}

This simulation investigates an FOSMC strategy applied to the PMSM system under the same conditions and model parameters as defined in the sections above. Fractional-order controllers generalize traditional PID and SMC structures by incorporating memory effects through non-integer-order integration and differentiation, thus offering improved tuning flexibility and smoother dynamics.

This study employs the Grünwald-Letnikov definition of a fractional derivative because it is well-suited for numerical applications and effectively captures behaviors that are nonlocal and dependent on memory \cite{tepljakov2013design}. The Grünwald-Letnikov fractional derivative is expressed as \cite{monje2010fractional}:

\begin{align}
    {}_{t_0}D_t^\alpha f(t) = \lim_{h \to 0} \frac{1}{h^\alpha} \sum_{j=0}^{\left\lfloor\frac{t-t_0}{h}\right\rfloor} (-1)^j \binom{\alpha}{j} f(t - jh)
\end{align}

In this expression:
\begin{itemize}
    \item $\alpha$: Fractional order of differentiation, extends the concept of traditional integer-order derivatives, allowing for non-integer values,
    \item $\binom{\alpha}{j}$: Generalized binomial coefficient, defined as:
    \[
    \binom{\alpha}{j} = \frac{\Gamma(\alpha+1)}{\Gamma(j+1)\Gamma(\alpha-j+1)},
    \]
    where $\Gamma(\cdot)$ represents the gamma function, which extends the concept of factorial to non-integer numbers,
    \item $f(t - jh)$: Function evaluated at discrete past time steps, weighted by $(-1)^j \binom{\alpha}{j}$,
    \item $h^\alpha$: Scaling factor based on the step size $h$.
\end{itemize}

The summation $\sum_{j=0}^{\left\lfloor\frac{t-t_0}{h}\right\rfloor}$ emphasizes the nonlocal characteristics of fractional derivatives by including historical function data points. This aspect makes the Grünwald-Letnikov definition especially suitable for modeling and control applications that require precise representation of dynamic systems with memory effects. Employing this definition guarantees the successful integration of fractional dynamics into the numerical implementation of the FOPID controller.

Following the formulation in \cite{zaihidee2019application}, the tracking error is defined as \(e = \omega_r^{\star} - \omega_r\), and the sliding surface incorporates fractional derivatives and integrals of the error:
\begin{equation}
s(t) = k_p e(t) + k_i\,\prescript{}{0}{\mathrm D}_t^{-\alpha} e(t)
       + k_d\,\prescript{}{0}{\mathrm D}_t^{\;\beta} e(t), \qquad 0 < \alpha, \beta < 1,
\end{equation}
Where \(k_p\), \(k_i\), and \(k_d\) are the proportional, integral, and derivative gains, respectively. The operators \(\prescript{}{0}{\mathrm D}_t^{-\alpha}\) and \(\prescript{}{0}{\mathrm D}_t^{\beta}\) denote the Grünwald–Letnikov fractional integral and derivative of orders \(\alpha\) and \(\beta\), respectively.

The reaching law is defined as that
\begin{equation}
\dot{s} = -w\,s - k_s\,\operatorname{sgn}(s), \qquad w, k_s > 0,
\end{equation}
which combines linear and discontinuous control actions to ensure finite-time convergence while mitigating steady-state chattering.

By substituting this dynamic into the mechanical torque balance of the PMSM, the torque-producing current \(i_q^{\star}(t)\) is derived as:
\begin{equation}
\begin{aligned}
i_q^{\star}(t) = \frac{1}{\chi} \Bigl[
    & \dot{\omega}_r^{\star}
    + w\,k_i\,\prescript{}{0}{\mathrm D}_t^{-\alpha} e
    + w\,k_d\,\prescript{}{0}{\mathrm D}_t^{\;\beta} e
    + k_s \int \operatorname{sgn}(s)\,dt \\
    &+ (w + a)\,k_p e \phantom{=} + k_i\,\prescript{}{0}{\mathrm D}_t^{1-\alpha} e
    + k_d\,\prescript{}{0}{\mathrm D}_t^{\beta + 1} e \Bigr]
\end{aligned}
\end{equation}
where
\begin{align*}
    \chi = \frac{3p\psi_f}{2J}, \quad a = \frac{B}{J}.
\end{align*}

The use of fractional operators enables improved control, smoother dynamics, and finer tuning of transient responses compared to integer-order SMC. Moreover, the combination of fractional integration and differentiation provides flexibility in shaping the sliding manifold response, enabling superior performance in both tracking and robustness to modeling uncertainties.

\begin{figure*}
  \hspace*{-0.18\textwidth}
  \begin{subfigure}[b]{0.8\textwidth}     
    \centering
    \includegraphics[width=\linewidth,trim=12bp 8.5cm 12bp 8.5cm, clip]
            {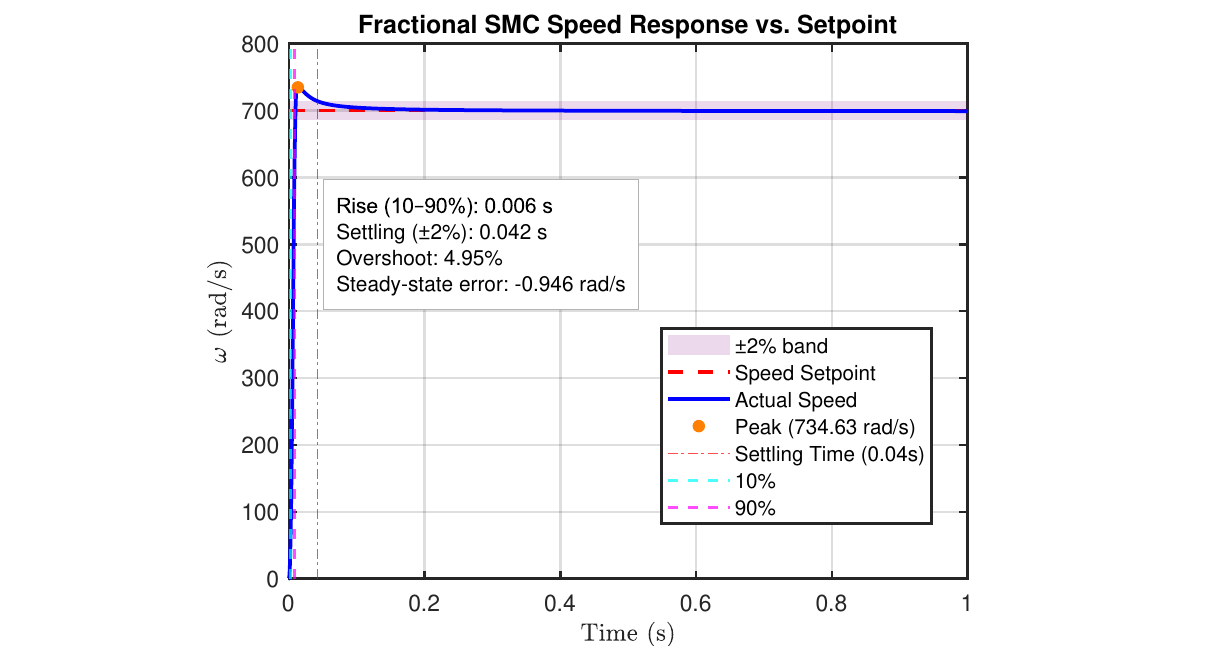}
    \caption{No external disturbance.}
    \label{fig:fsmc_nodist}
  \end{subfigure}%
  \hspace*{-0.25\textwidth}                 
  \begin{subfigure}[b]{0.8\textwidth}
    \centering
    \includegraphics[width=\linewidth,trim=12bp 8.5cm 12bp 8.5cm, clip]
            {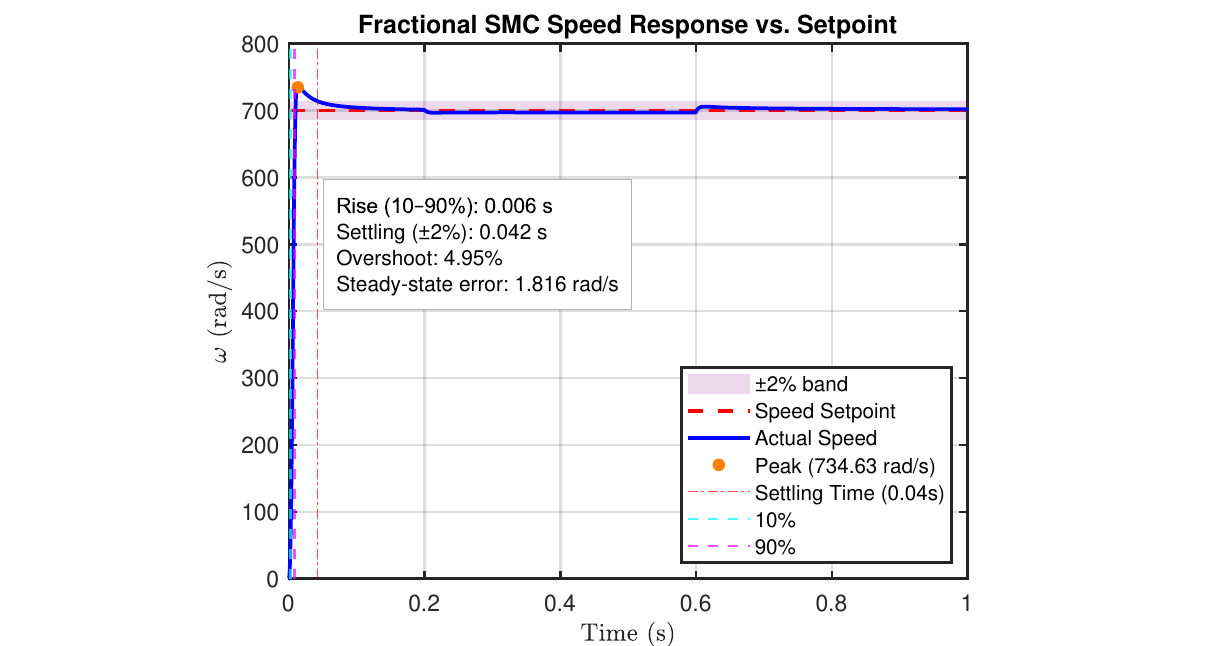}
    \caption{External disturbance applied at \(t=0.2\,\text{s}\) and \(t=0.6\,\text{s}\).}
    \label{fig:fsmc_dist}
  \end{subfigure}
  \caption{Rotor-speed response under fractional order sliding mode control.}
  \label{fig:fsmc_pair}
\end{figure*}

Figures \ref{fig:fsmc_nodist} and \ref{fig:fsmc_dist} represent the speed response of the FOSMC.
For a single-step command, FOSMC achieves fast tracking with modest overshoot. Without disturbance, the speed response attains a rise time of \(0.0058\,\text{s}\), a settling time (2\% band) of \(0.0448\,\text{s}\), an overshoot of \(5.09\%\), and \(\approx 0\) a steady-state error. With disturbance, rise and settling remain essentially unchanged \((0.0058\,\text{s},\,0.0448\,\text{s})\) and the global overshoot is similar, while a slight bias appears at steady state \((e_{ss}\approx -2.76\) on a \(\sim 699\) setpoint, \(\approx 0.40\%\)). Around the disturbance instants, the excursions stay well inside the \(\pm2\%\) band: near \(t=0.2\,\text{s}\), \(\max\{+3.15,-2.39\}\) (\(\approx0.45\%\), \(0.34\%\)); near \(t=0.6\,\text{s}\), \(\max\{+6.72,-1.98\}\) (\(\approx0.96\%\), \(0.28\%\)). Overall, FOSMC preserves sub-\(50\) ms settling and band-limited regulation under disturbances; minor tuning of the fractional order and boundary layer could reduce the initial \(5\%\) overshoot and the slight steady-state bias under load.

\subsection{Super-Twisting Sliding-Mode Control (STSMC)}

To mitigate chattering while preserving the robustness properties of sliding mode control, a second-order STSMC algorithm is applied in this simulation. The PMSM plant and simulation settings are identical to those used in previous sections.

The STSMC control law can be defined by the equation below, which consists of the equivalent, continuous, and discontinuous parts \cite{heng2017design}.
\begin{equation}
  u(t) = u_{eq}(t)\;-\;\operatorname{sgn}(g)\,L\,|s(t)|^{1/2}\,\operatorname{sgn}\!\big(s(t)\big)\;+\;u_1(t),
\end{equation}
\begin{equation}
  \dot u_1(t) = -\,\operatorname{sgn}(g)\,W\,\operatorname{sgn}\!\big(s(t)\big).
\end{equation}

\noindent where \(L,W>0\) are gains and \(u_1(t)\) is an auxiliary state. The equivalent term
\(u_{eq}\) is obtained by equation  \(\dot s\) to $0$. The equivalent control law can thus be derived as follows:
\begin{equation}
    \dot{s} = c\dot{e} \, + \ddot{e} = 0
\end{equation}
Taking the derivation of \ref{eq:error} and substituting \ref{eq:mechanical}, the following equation can be obtained ;

\begin{equation}
\dot{s} = f(x,t) + g \dot{i}_q, \qquad g = -\frac{k_t}{J},
\tag{5}
\end{equation}
where
\begin{equation}
\begin{aligned}
f(x,t) &= \ddot{\omega}_{\mathrm{ref}} + c\,\dot{\omega}_{\mathrm{ref}}
        + \left( \frac{B k_t}{J^2} - \frac{c k_t}{J} \right) i_q
        + \left( -\frac{B^2}{J^2} + \frac{c B}{J} \right) \omega_m \\
&\quad + \left( -\frac{B}{J^2} + \frac{c}{J} \right) T_L
        + \frac{\dot{T}_L}{J}.
\end{aligned}
\tag{6}
\end{equation}
when
\begin{equation}
  u_{eq} \triangleq \dot{i}_q , \quad
  u_{eq} = -\frac{f}{g} = \frac{J}{k_t} f(x,t).
\end{equation}
Thus
\begin{equation}
\begin{aligned}
u_{eq}
&= \frac{J}{k_t}\bigl(\ddot{\omega}_{\mathrm{ref}} + c\,\dot{\omega}_{\mathrm{ref}}\bigr)
  + \left(\frac{B}{J} - c\right) i_q
  + \frac{B}{k_t}\left(c - \frac{B}{J}\right)\omega_m \\
&\quad + \frac{1}{k_t}\left(c - \frac{B}{J}\right) T_L
  + \frac{1}{k_t}\dot{T}_L .
\end{aligned}
\end{equation}

The STSMC structure provides finite-time convergence to the sliding manifold , while significantly reducing chattering through its continuous control action. Simulation results demonstrate smooth transient performance and robustness to disturbances when compared to conventional first-order SMC.

\begin{figure*}[htbp]
    \hspace*{-0.18\textwidth}
  \begin{subfigure}[b]{0.8\textwidth}      
    \centering
    
   \includegraphics[ width=\linewidth, trim=12bp 8.5cm 12bp 8.5cm, clip]
            {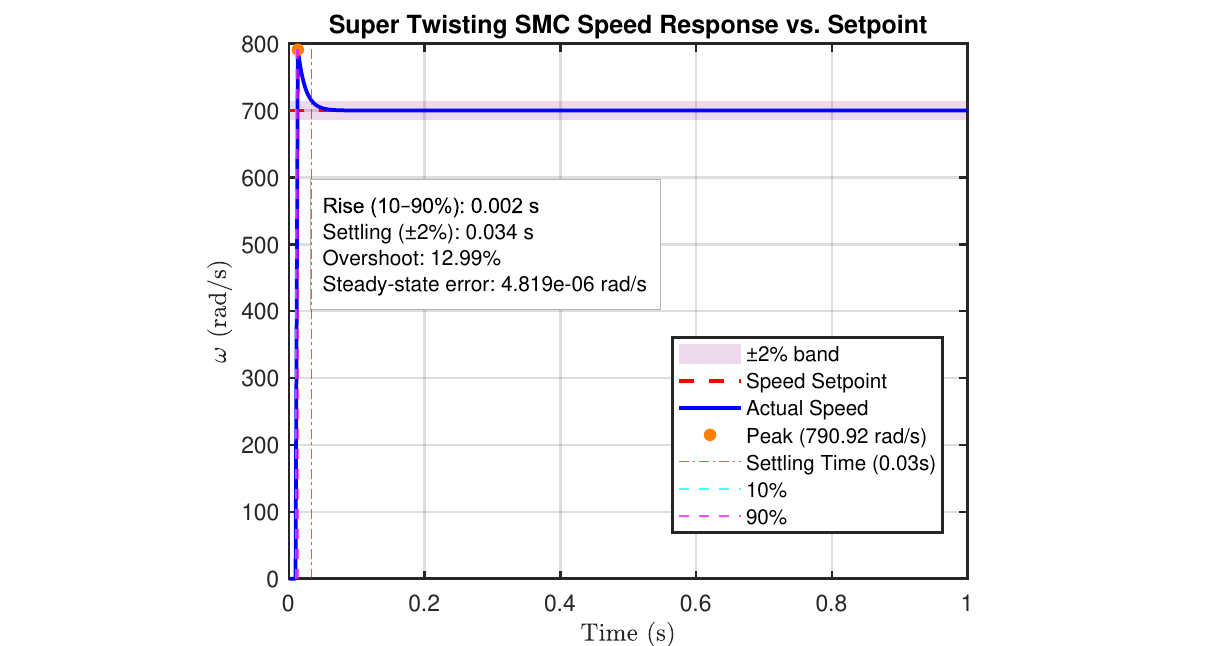}
    \caption{No external disturbance.}
    \label{fig:stsmc_nodist}
  \end{subfigure}%
  \hspace*{-0.25\textwidth}                 
  \begin{subfigure}[b]{0.8\textwidth}
    \centering
   \includegraphics[ width=\linewidth, trim=12bp 8.5cm 12bp 8.5cm, clip]
            {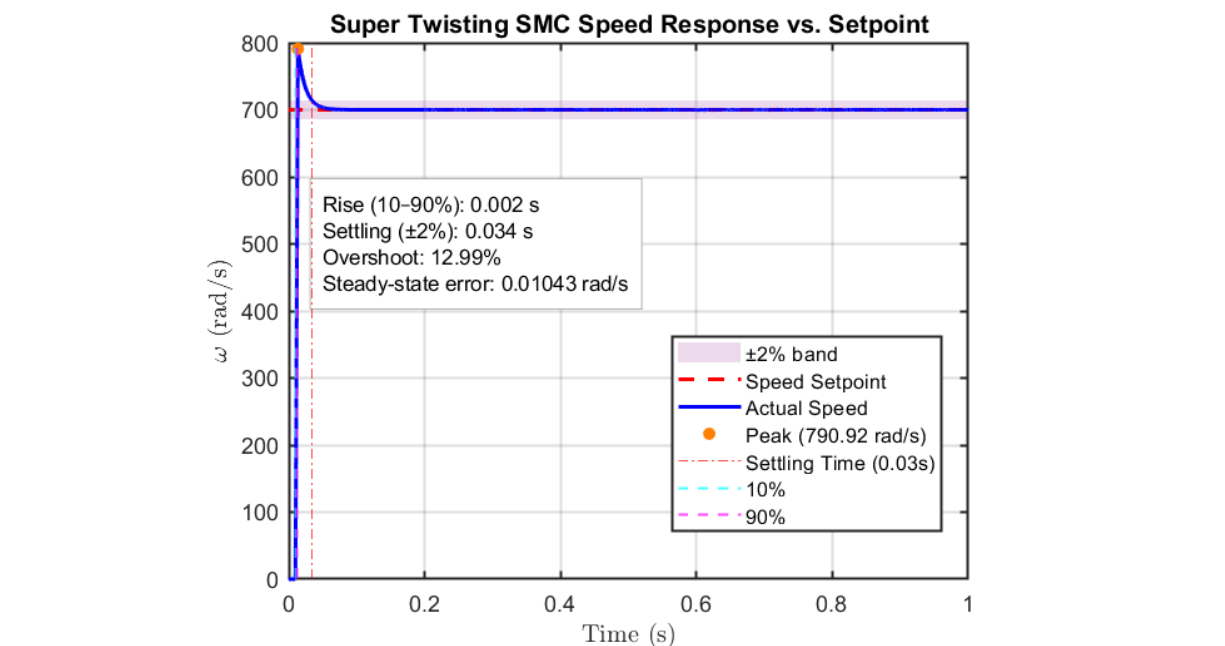}
    \caption{External disturbance applied at \(t=0.2\) s and \(t=0.6\) s..}
    \label{fig:stsmc_dist}
  \end{subfigure}
    \caption{Rotor-speed response under super-twisting SMC..}
  \label{fig:stsmc_pair}
\end{figure*}

Figures \ref{fig:stsmc_nodist} and \ref{fig:stsmc_dist} illustrate the speed response of the STSMC. Under a single-step command, STSMC yields high-speed transients with effective disturbance rejection. In the disturbance–free case, the speed response achieves a rise time of \(0.00239\,\text{s}\), a settling time (2\% band) of \(0.02423\,\text{s}\), and an overshoot of \(12.99\%\), with \(\approx 0\) steady–state error. With disturbance, rise and settling remain essentially unchanged \((0.00239\,\text{s},\,0.02423\,\text{s})\) and the steady–state error remains negligible. Overall, STSMC preserves millisecond–scale convergence and zero steady–state error in the presence of disturbances; the main trade-off is a higher peak overshoot compared to integral/terminal variants, which could be mitigated via tuning of the super–twisting gains and boundary-layer thickness.

\section{Analysis and Discussion}
\begin{figure*}[htbp]
     \hspace*{-0.18\textwidth}
  \begin{subfigure}[b]{0.8\textwidth}     
    \centering
    \includegraphics[ width=\linewidth, trim=12bp 8.5cm 12bp 8.5cm, clip]{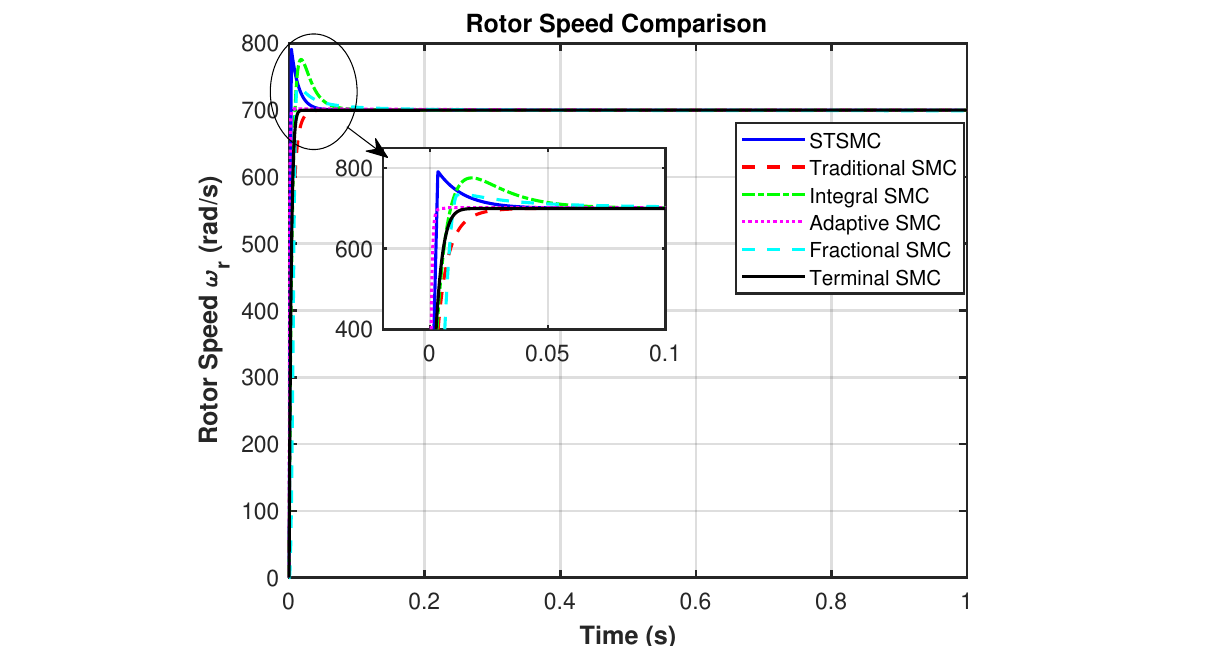}
    \caption{No external disturbance.}
    \label{fig:joint_nodist}
  \end{subfigure}
  \hspace*{-0.25\textwidth}                 
  \begin{subfigure}[b]{0.8\textwidth}
    \centering
   \includegraphics[ width=\linewidth, trim=12bp 8.5cm 12bp 8.5cm, clip]{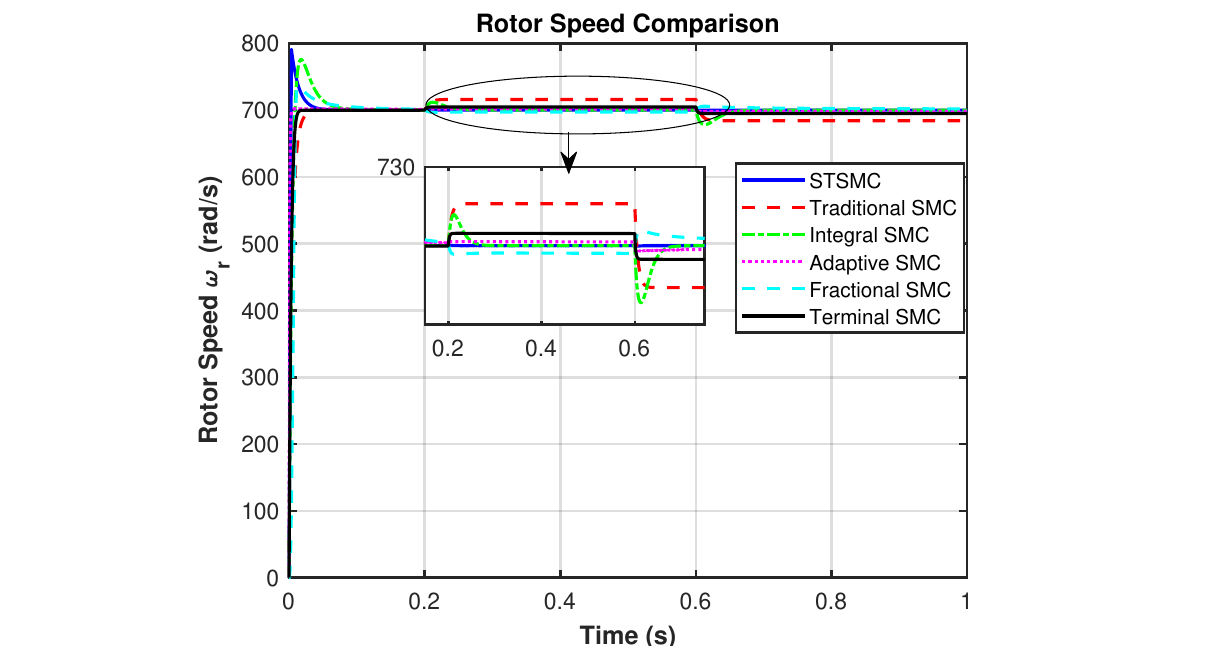}
    \caption{External disturbance applied at \(t=0.2\) s and \(t=0.6\) s.}
  \end{subfigure}
  \caption{Comparison of rotor speed response.}
  \label{fig:joint_pair}
\end{figure*}

Figure~\ref{fig:joint_pair} presents a consolidated comparison of the rotor speed trajectories for all six SMC strategies evaluated in this study, CSMC, FOSMC, ISMC, STSMC, TSMC, and ASMC. Two scenarios are considered: (a) nominal step input with no external disturbances and (b) identical input with load torque disturbances applied at \(t = 0.2\ \text{s}\) and \(t = 0.6\ \text{s}\). 
Across the six SMC variants, conventional, integral, terminal, adaptive, fractional–order, and super–twisting—the trade–offs between speed, overshoot, and disturbance rejection are clear. TSMC delivers the best all–around transient under a single step: millisecond–scale rise (\(\sim\! 6.2\,\text{ms}\)), the fastest settling among SMCs (\(\sim\! 10\,\text{ms}\)), and negligible overshoot without disturbance (\(\approx 0\%\), rising to only \(\sim\!1.44\%\) under disturbance) while maintaining zero steady–state error and band–limited excursions (\(<1\%\)). ISMC guarantees zero steady–state error in both scenarios but pays with a much longer settling under disturbance (\(\sim\! 0.63\,\text{s}\); \(\approx 12.4\times\) slower than no–disturbance), despite a moderate overshoot (\(\sim\!10.8\%\)) and a fast rise (\(\sim\! 6.2\,\text{ms}\)). STSMC achieves the quickest rise (\(\sim\! 2.4\,\text{ms}\)) with robust disturbance rejection and near–zero bias, but at the cost of the highest overshoot (\(\sim\!13\%\)) and a settling of \(\sim\! 24\,\text{ms}\). FOSMC provides a middle ground—rise \(\sim\!5.8\,\text{ms}\), settling \(\sim\!44.8\,\text{ms}\), overshoot \(\sim\!5.1\%\) exhibiting only a small steady–state bias (\(\approx 0.4\%\)) under disturbance. In contrast, CSMC is fast and accurate without disturbance (rise \(\sim\!10.3\,\text{ms}\), settle \(\sim\!20.4\,\text{ms}\), near–zero overshoot) but develops a persistent \(\sim\!2.3\%\) bias with disturbance and does not re–enter the \(2\%\) band within the window.

\subsection{Step Response with Load Disturbance} \label{sec:quant_results_dist}

Table~\ref{tab:perf_idx_dist} summarizes the integral performance indices—ISE, IAE, ITSE, and ITAE—under the presence of torque disturbances. Boldface entries denote the best performance for each metric.

\begin{table*}[htbp]
  \centering
  \caption{Integral performance indices (ISE, IAE, ITSE, ITAE) under nominal and disturbed conditions.}
  \label{tab:perf_idx_side_by_side}
  
  \begin{subtable}[htbp]{0.48\textwidth}
    \centering
    \caption{Without external disturbance}
    \label{tab:perf_idx_nodist}
    \begin{tabular}{lcccc}
      \toprule
      Controller & ISE & IAE & ITSE & ITAE \\
      \midrule
      Conventional   & 1015.72 &  3.15 &  2.30 & 0.02 \\
      Fractional     &  235.43 & 11.26 & 60.38 & 5.47 \\
      Terminal       &  749.25 &  2.09 &  1.10 & \textbf{0.012} \\
      Integral       &  947.00 &  4.22 &  3.73 & 0.070 \\
      Adaptive       &  149.02 & \textbf{0.85} & \textbf{0.106} & 0.067 \\
      Super-twisting & \textbf{125.60} & 1.32 & 0.632 & 0.432 \\
      \bottomrule
    \end{tabular}
  \end{subtable}
  \hfill
  \begin{subtable}[htbp]{0.48\textwidth}
    \centering
    \caption{With external disturbance}
    \label{tab:perf_idx_dist}
    \begin{tabular}{lcccc}
      \toprule
      Controller & ISE & IAE & ITSE & ITAE \\
      \midrule
      Conventional   & 1214.01 & 15.67 & 122.20 & 7.57 \\
      Fractional     & 235.43  & 11.26 &  60.38 & 5.47 \\
      Terminal       & 768.74  &  6.01 &  13.30 & 2.42 \\
      Integral       & 960.88  &  5.30 &  11.15 & \textbf{0.60} \\
      Adaptive       & 150.79  &  1.93 &  1.21  & 0.76 \\
      Super-twisting &  \textbf{125.60} & \textbf{1.32} & \textbf{0.63} & 0.43 \\
      \bottomrule
    \end{tabular}
  \end{subtable}
\end{table*}

Under load disturbance, Super-Twisting SMC (STSMC) consistently achieves the lowest ISE and ITSE values, confirming its strong robustness and finite-time convergence. Adaptive SMC (ASMC) closely follows, offering the lowest IAE and minimal overshoot (0.28\%). While Fractional SMC (FOSMC) settles rapidly (1.5 ms), it exhibits a steady-state error due to limited gain tuning. Integral SMC (ISMC) entirely eliminates steady-state bias but suffers from excessive overshoot and delayed settling—conventional and Terminal SMC lag in both transient and steady-state performance.

In summary, STSMC remains the most resilient under disturbance, while ASMC offers a highly competitive tradeoff between precision and smoothness. FOSMC is favorable in fast-response scenarios where a small residual error is acceptable.

\subsection{Control Response}

\begin{figure*}
    \centering
    \includegraphics[width=\linewidth, keepaspectratio]{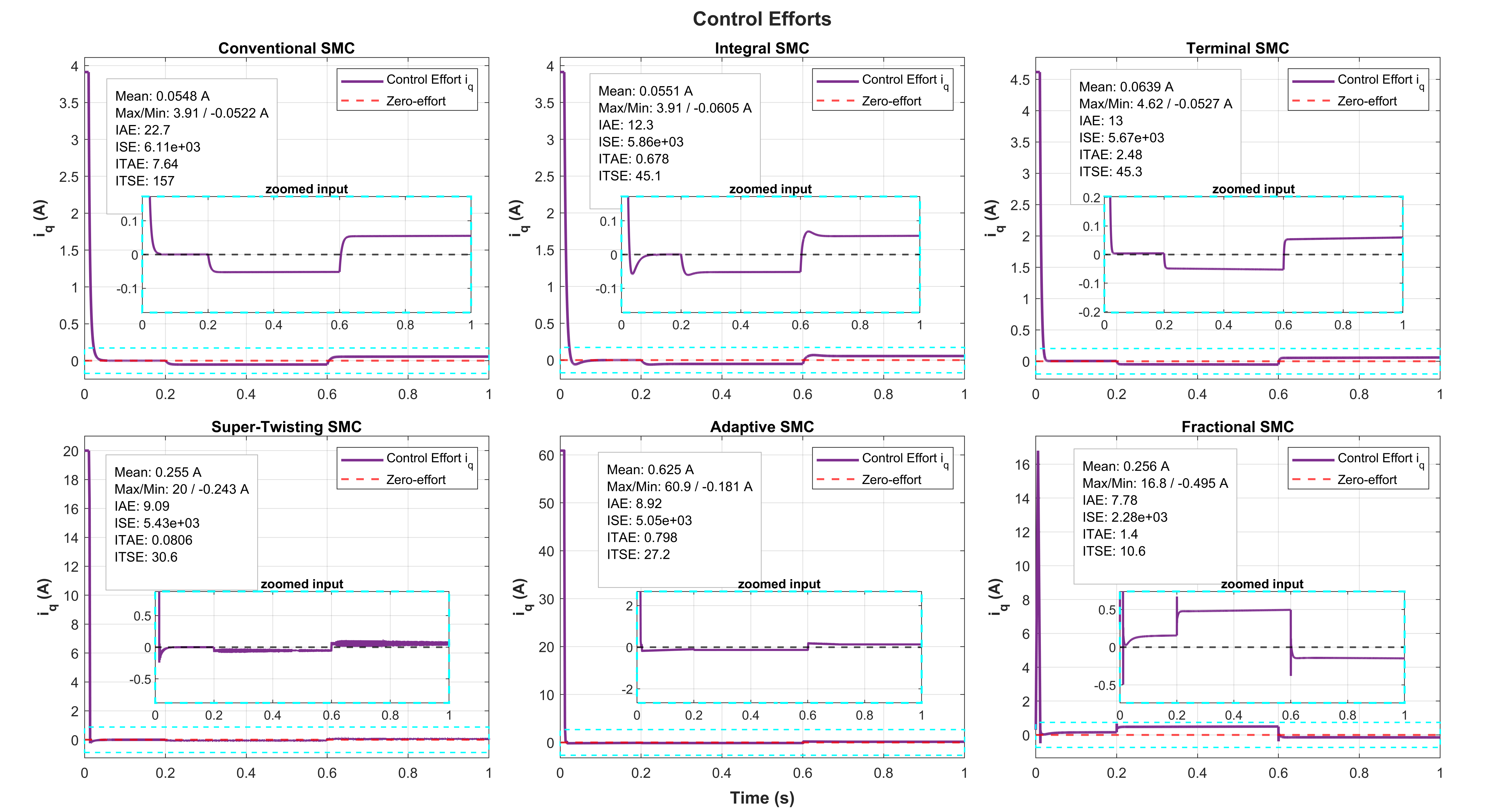}
    \caption{Control Effort with disturbance}
    \label{fig: placeholder}
\end{figure*}

\begin{figure*}
    \centering
    \includegraphics[width=\linewidth, keepaspectratio]{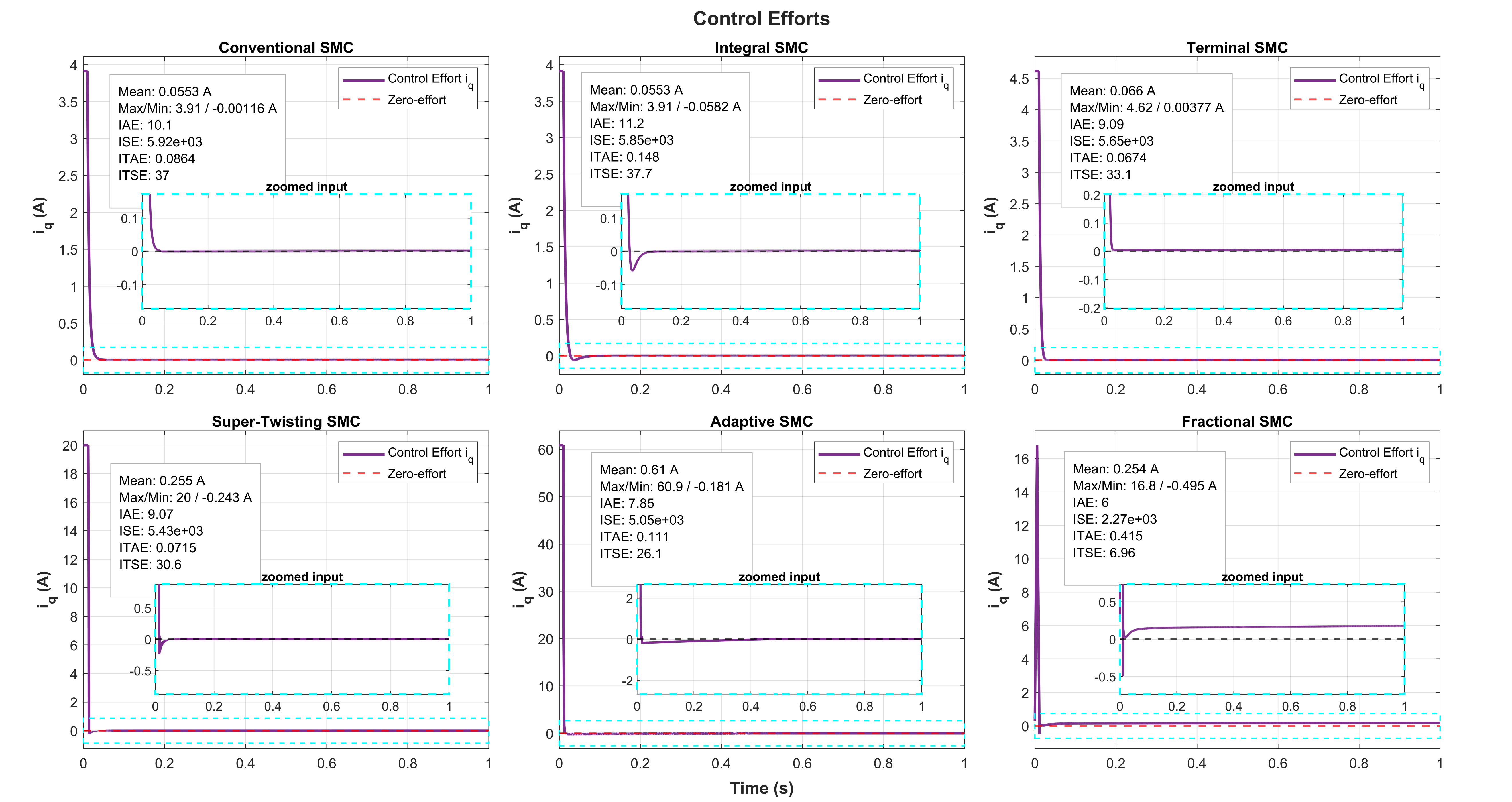}
    \caption{Control Effort with no disturbance}
    \label{fig: placeholder2}
\end{figure*}

\begin{figure*}
     \hspace*{-0.18\textwidth}
  \begin{subfigure}[b]{0.8\textwidth}     
    \centering
    \includegraphics[ width=\linewidth, trim=12bp 8.5cm 12bp 8.5cm, clip]{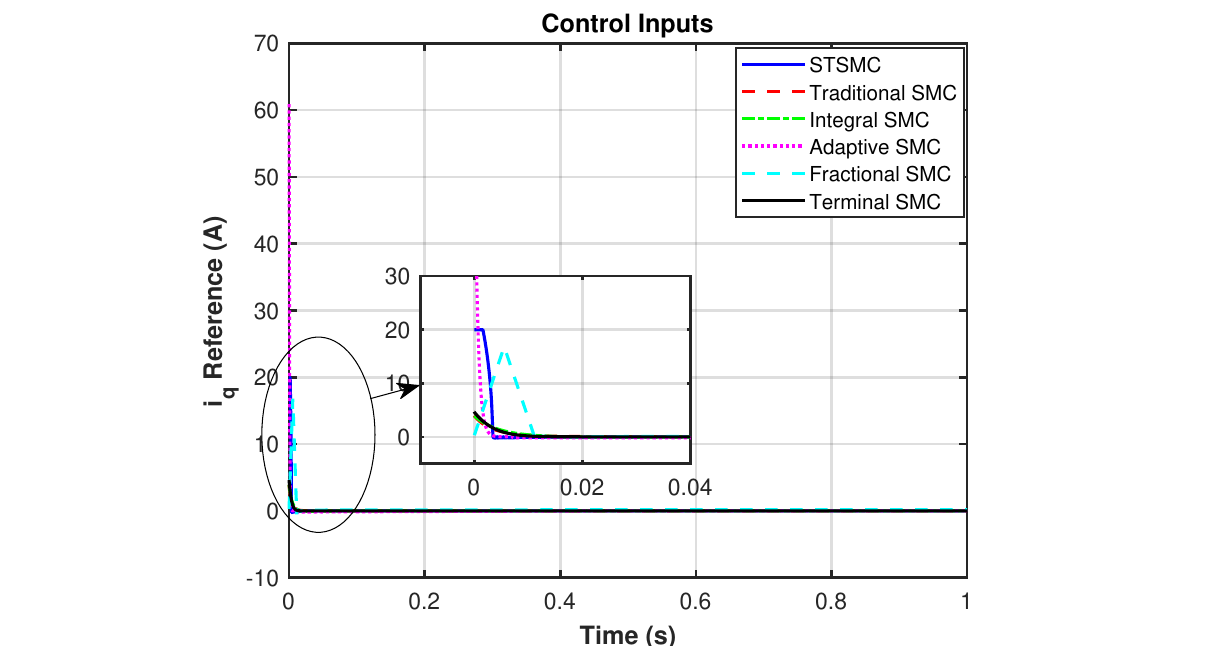}
    \caption{No external disturbance.}
    \label{fig:joint_contr_nodist}
  \end{subfigure}
  \hspace*{-0.25\textwidth}                 
  \begin{subfigure}[b]{0.8\textwidth}
    \centering
   \includegraphics[ width=\linewidth,trim=12bp 8.5cm 12bp 8.5cm, clip]{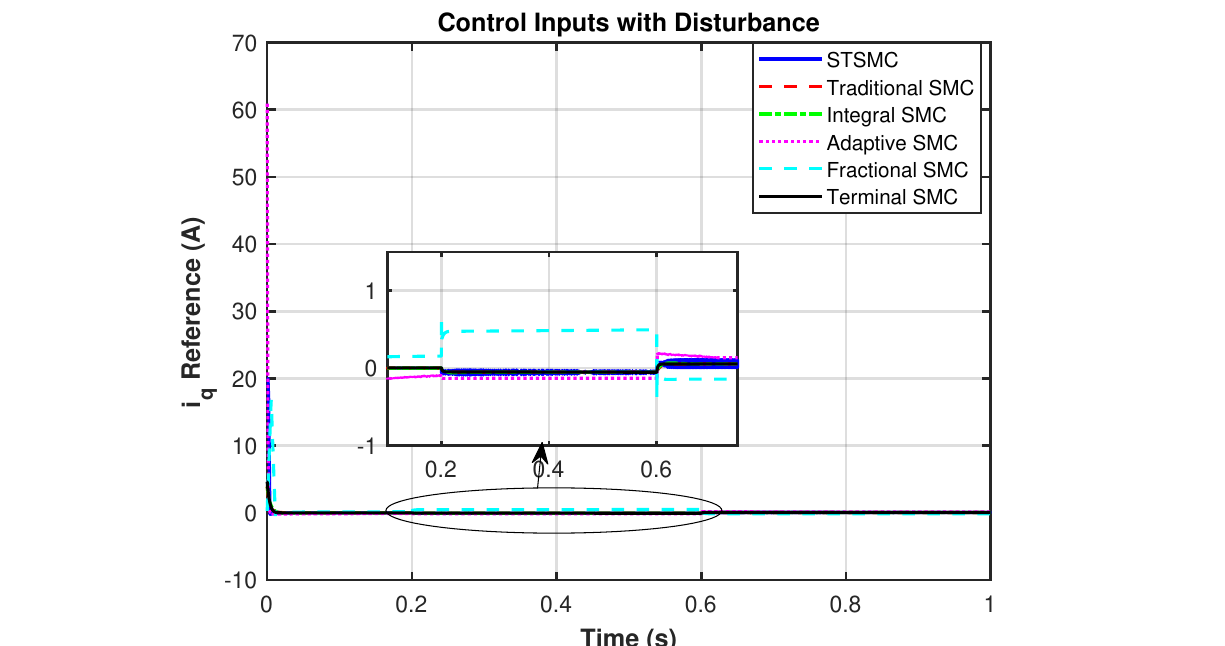}
    \caption{External disturbance applied at \(t=0.2\) s and \(t=0.6\) s.}
  \end{subfigure}
  \caption{Comparison of rotor speed response.}
  \label{fig:joint_contr_pair}
\end{figure*}

\subsection{Step Response \emph{without} Load Disturbance}\label{sec:quant_results_nodist}

In the nominal case, the PMSM is commanded from 0 to \(700\ \text{r/min}\) without external torque perturbations. Table~\ref{tab:perf_idx_nodist} lists the resulting integral error metrics.

STSMC again shows the lowest ISE and second-lowest IAE, validating its energy-efficient convergence. ASMC exhibits the best values in IAE, ITSE, and ITAE, indicating a rapid and smooth settling. FOSMC achieves the fastest rise time but retains a steady-state bias due to reduced switching activity. While Terminal and Conventional SMC achieve zero overshoot, they require longer settling times. ISMC effectively eliminates steady-state error, but this comes at the cost of high initial overshoot. Overall, STSMC and ASMC stand out as the most balanced options when both speed and precision are prioritized, making them strong candidates for high-performance PMSM applications.

\subsection{Comparative Interpretation of SMC Variants}

The six investigated SMC variants exhibit distinct performance behaviors arising from their underlying dynamic structures. 
Conventional SMC enforces robustness through a discontinuous switching law, providing rapid convergence under nominal conditions but leaving a steady-state bias when load disturbances appear. 
Its limited disturbance rejection stems from the absence of integral or higher-order compensation. 

\textbf{Integral SMC (ISMC)} eliminates this bias by initializing system trajectories on the sliding manifold; the integral term maintains invariance to matched disturbances. 
However, the same memory action lengthens the transient and increases settling time after sudden torque changes. 

\textbf{Terminal SMC (TSMC)} introduces nonlinear manifold dynamics that accelerate convergence as the error decreases. 
The nonlinear gain compresses the transient within a finite time, ensuring rapid disturbance recovery with near-zero steady-state error. 

\textbf{Fractional-Order SMC (FOSMC)} enhances tuning flexibility by embedding non-integer differentiation and integration. 
The fractional operators endow the controller with a ``memory'' effect, improving smoothness and short-term tracking but leaving a slight residual bias when the fractional orders are not perfectly matched to system inertia. 

\textbf{Adaptive SMC (ASMC)} automatically modulates the switching gain according to error magnitude, strengthening control effort far from the surface and reducing it near equilibrium. 
This adaptive mechanism yields low chattering and excellent transient regulation without requiring prior knowledge of disturbance bounds. 

Finally, the \textbf{Super-Twisting SMC (STSMC)}, a second-order formulation, provides continuous control action by integrating the switching term. 
This smooth actuation suppresses chattering while maintaining finite-time convergence, explaining its consistently superior energy-error indices in both nominal and disturbed cases.

\subsection{Quantitative Performance Analysis}

The integral metrics (\emph{ISE, IAE, ITSE, ITAE}) in Table 1 quantify the trade-offs among these designs. 
\begin{itemize}
  \item \textbf{STSMC} achieved the lowest ISE and ITSE in both scenarios, confirming that its continuous second-order dynamics minimize total error energy and yield the most disturbance-resilient behavior. 
  \item \textbf{ASMC} provided the smallest IAE and ITAE, indicating smoother error decay and minimal overshoot due to online gain adaptation. 
  \item \textbf{FOSMC} ranked close in rise and settling times but retained a minor steady-state offset because its fractional-order memory filters the discontinuous action. 
  \item \textbf{ISMC} achieved perfect steady-state accuracy yet required the longest recovery after disturbances, revealing that robustness from the integral term is gained at the expense of dynamic agility. 
  \item \textbf{TSMC} offered fast finite-time convergence and strong transient performance, while \textbf{Conventional SMC} served as the baseline—simple and robust but less precise under model uncertainty.
\end{itemize}

The combined comparison clarifies that higher-order and adaptive mechanisms outperform first-order designs not merely in smoothness but also in energy efficiency: both STSMC and ASMC maintain small control amplitudes once sliding is reached, reducing equivalent chattering power.

\subsection{Implementation and Practical Insights}

From an implementation standpoint, the controllers differ markedly in complexity and computational demand. 
\begin{itemize}
  \item \textbf{Conventional SMC} and \textbf{TSMC} require minimal computation and are suitable for low-cost digital drives where moderate ripple is acceptable. 
  \item \textbf{ISMC} demands integral state storage but remains lightweight for fixed-point DSPs. 
  \item \textbf{FOSMC} introduces fractional operators whose numerical approximation increases memory and CPU load, restricting its use to high-speed microcontrollers or offline tuning studies. 
  \item \textbf{ASMC} and \textbf{STSMC} involve adaptive or auxiliary-state dynamics but can be efficiently implemented with discrete-time algorithms and modest sampling rates. 
\end{itemize}

For embedded real-time PMSM drives, the STSMC-ASMC family offers the best balance between computational feasibility and control precision. In contrast, FOSMC is better suited for research and hardware-in-the-loop evaluation.

\subsection{Design Trade-offs and Research Implications}
Three practical trade-offs summarize the benchmark findings: robustness–smoothness, convergence–computation, and generality–tuning effort. These jointly define the design space for future PMSM sliding-mode controllers, guiding researchers toward an optimal balance between real-time feasibility and dynamic precision.
\begin{enumerate}
  \item \textbf{Robustness vs.~Smoothness:} Higher-order (STSMC) and adaptive (ASMC) designs achieve smooth actuation without sacrificing robustness, while conventional and terminal laws retain small residual ripple. 
  \item \textbf{Convergence vs.~Computation:} Fractional and adaptive schemes accelerate convergence but impose higher computational costs. 
  \item \textbf{Generality vs.~Tuning Effort:} Integral and adaptive variants exhibit broader disturbance tolerance but require careful parameter initialization to avoid slow recovery.
\end{enumerate}

These findings suggest several research directions:
\begin{itemize}
  \item Development of \textbf{adaptive fractional-order SMC} that fuses memory and gain-scheduling features to combine fast response with bias elimination. 
  \item Exploration of \textbf{multi-machine and fault-tolerant} configurations using coordinated STSMC or ISMC frameworks. 
  \item Real-time \textbf{hardware-in-loop benchmarking} to assess computational overhead and switching losses. 
  \item Integration of \textbf{learning-based optimization} (e.g., reinforcement or evolutionary tuning) for online adjustment of reaching laws.
\end{itemize}

\subsection{Algorithmic and Computational Complexity Analysis}
Table~\ref{tab: complexity} summarizes the relative computational effort of each SMC variant per sampling period, considering arithmetic operations, stored states, and auxiliary dynamics.  
Among the baseline schemes, CSMC and ISMC exhibit the lowest complexity, while TSMC introduces minor computational overhead due to additional switching logic.  
STSMC and ASMC moderately increase the computational load through auxiliary differential and adaptive-gain updates, respectively; yet, both remain feasible within a $100-\mu$ sampling period on a $100 MHz$ DSP.  In contrast, FOSMC incurs the highest cost owing to the discrete approximation of fractional-order operators, which demands numerous buffer updates per iteration.

\begin{table}[t]
\centering
\caption{Algorithmic complexity comparison of SMC variants per sampling period.}
\label{tab: complexity}
\resizebox{\columnwidth}{!}{
\begin{tabular}{lccc}
\toprule
Controller & Extra States & Multiplications\footnotemark & Memory \\
\midrule
CSMC        & 0                    & $\sim$12   & Low       \\
ISMC        & 1 (integral)         & $\sim$15   & Low–Med   \\
TSMC        & 0                    & $\sim$18   & Low      \\
FOSMC       & 2 (FO buffers)       & $>50$      & High      \\
ASMC        & 1 (adaptive gain)    & $\sim$25   & Medium    \\
STSMC       & 1 (aux.\ $\eta$)     & $\sim$22   & Medium  \\
\bottomrule
\end{tabular}
}
\end{table}
\footnotetext{Operation counts are approximate and depend on specific discretization and implementation.}


\begin{table}[t]
\centering
\caption{Relative computational complexity of SMC variants (normalized to CSMC).}
\label{tab:complexity_bar}
\begin{tabular}{@{}ll@{}}
\toprule
\textbf{Controller} & \textbf{Relative Cost (visualized)} \\
\midrule
CSMC & \rule{1.0cm}{2pt}~(1.0$\times$) \\
ISMC & \rule{1.2cm}{2pt}~(1.2$\times$) \\
TSMC & \rule{1.3cm}{2pt}~(1.3$\times$) \\
ASMC & \rule{1.6cm}{2pt}~(1.6$\times$) \\
STSMC & \rule{1.5cm}{2pt}~(1.5$\times$) \\
FOSMC & \rule{3.0cm}{2pt}~(3.0$\times$) \\
\bottomrule
\end{tabular}
\vspace{1mm}
\begin{minipage}{0.9\linewidth}
\footnotesize
\textit{Note:} Bar length indicates relative computational effort per control update. Values are normalized with respect to CSMC baseline complexity. FOSMC exhibits the highest computational load due to fractional derivative operations.
\end{minipage}
\end{table}

\section{Conclusion} \label{sec:conclusion}
This study established a unified benchmarking framework for evaluating six major Sliding-Mode Control (SMC) strategies for PMSM speed regulation. Under identical modeling, test, and evaluation conditions, the benchmark quantitatively compared conventional, integral, terminal, fractional-order, adaptive, and super-twisting controllers. Among the evaluated methods, the Super-Twisting and Adaptive SMCs achieved the best compromise between robustness, smoothness, and computational efficiency. At the same time, the fractional-order approach offered the fastest transient response with negligible steady-state bias. Beyond performance analysis, the framework incorporated a structured parameter-tuning guideline and quantified computational complexity, offering practical insight into real-time controller selection. Overall, this benchmark provides a reproducible foundation for extending SMC design toward adaptive-fractional, observer-based, and learning-enhanced control paradigms in advanced PMSM drive applications.

\begin{table}[htbp]
   \centering
   \caption{List of acronyms used throughout the paper for key motor-control and optimization strategies.}
   \label{tab:acronyms}
   \renewcommand{\arraystretch}{1.2}
   \begin{tabular}{|m{6em}|p{6.5cm}|}
   \hline
   \textbf{Acronym} & \textbf{Meaning} \\ 
   \hline
   \multicolumn{2}{|c|}{\textbf{Motor Type}} \\ 
   \hline
   PMSM & Permanent-Magnet Synchronous Motor \\ 
   \hline
   \multicolumn{2}{|c|}{\textbf{Sliding-Mode Control Variants}} \\ 
   \hline
   SMC   & Sliding-Mode Control \\ 
   ISMC  & Integral Sliding-Mode Control \\ 
   ASMC  & Adaptive Sliding-Mode Control \\ 
   TSMC  & Terminal Sliding-Mode Control \\ 
   NTSMC & Non-Singular Terminal Sliding-Mode Control \\ 
   FTSMC & Fast Terminal Sliding-Mode Control \\ 
   STSMC & Super-Twisting Sliding-Mode Control \\ 
   STA   & Super-Twisting Algorithm \\ 
   HOSMC & Higher-Order Sliding-Mode Control \\ 
   FOSMC & Fractional-Order Sliding-Mode Control \\ 
   SOSMC & Second-Order Sliding-Mode Control \\ 
   \hline
   \multicolumn{2}{|c|}{\textbf{Control Enhancements and Metrics}} \\ 
   \hline
   HRL   & Hybrid Reaching Law \\ 
   SMRL  & Sliding-Mode Reaching Law \\ 
   IAE   & Integral of Absolute Error \\ 
   ISE   & Integral of Squared Error \\ 
   ITAE  & Integral of Time-Weighted Absolute Error \\ 
   ITSE  & Integral of Time-Weighted Squared Error \\ 
   \hline
   \end{tabular}
\end{table}

\section*{Acknowledgments}

\bibliographystyle{elsarticle-num} 
\bibliography{references}
\end{document}